# A closed-form approximate incoherent GN-model supporting MCI contributions

**Mahdi Ranjbar Zefreh, Pierluigi Poggiolini**

*Politecnico di Torino*

***Abstract:*** In this paper, a closed form formula for nonlinearity modeling in modern coherent fiber optic communication systems is derived based on the incoherent GN model. The model covers multi-channel interference (MCI). The effectiveness of the derived formula is shown particularly in near zero dispersion environment. Finally, accuracy of the derived formula is improved by adding correction factors based on a large data test set.

## 1. Introduction

Real-time physical-layer-aware control and optimization of ultra-high-capacity optical networks is becoming an increasingly important aspect of networking, as throughput demand increases. To achieve it, accurate models of fiber non-linear effects (or NLI, Non-Linear-Interference) are needed, which must also be computable in real-time. While several effective NLI models are available [1]-[7], none of them is real-time in their native form, as they all include numerical integrals. Recently, though, Closed-Form Model (CFM) approximations of the GN/EGN models have been proposed [8],[9], capable of assessing whole links in fractions of a second. In [10] one such CFM was upgraded and tested over 7,000 highly randomized system scenarios, showing very good accuracy in reproducing the full-fledged numerically-integrated EGN-model, while being many orders of magnitude faster.

One limitation of the CFM [10] was however an increasing discrepancy vs. the EGN-model towards low fiber dispersion values, and especially for $D < 1.5\ ps/(nm.km)$.

The reason why this issue is significant is that, while most cables are based on SMF and operate at high dispersion, a considerable portion of deployed cables still hosts non-zero dispersion-shifted fibers (NZ-DSF), and even fibers whose dispersion zero is in the C-band (DSFs). With the impeding saturation of fiber bandwidth, there is currently a strong push towards using all available deployed fibers, including these low-dispersion fiber types. In addition, a further trend is towards using extended or alternative fiber bands which can be close or include a dispersion zero. As a result, in the upcoming bandwidth-constrained scenario, real-time NLI models for physical layer-aware optical networks must be able to handle near-zero or zero-dispersion fibers as well.

In this paper we analytically augment the real-time CFM [10] to make it capable of handling such environments. Finally, we test it both at low-to-zero dispersion, and over an enlarged version of the test-set used in [10], for a total of over 9,000 link configurations.

## 2. GN model formula

In general, the power spectral density (PSD) of each channel in WDM comb has raised cosine shape function with respect to frequency while for keeping simplicity we assume all the channels in WDM comb have rectangular shape functions with respect to frequency. In fact, if the channel is raised-cosine with nonzero roll-off, we approximately replace it with a rectangular channel with the same center frequency as the original raised cosine channel. Also, we consider the null-to-null bandwidth of the approximated rectangular channel equal to the symbol rate (baud rate) of the original raised cosine channel while keeping the constant value of the PSD of the rectangle function the same as the maximum value of the PSD of the original raised cosine channel as it is shown in Figure (1).

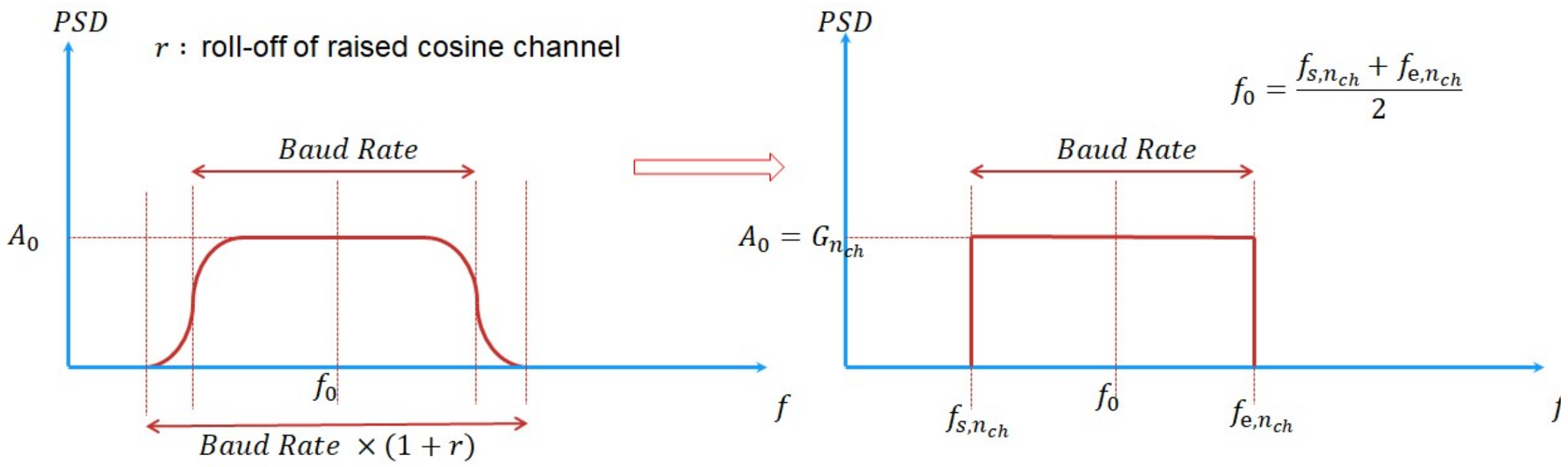

***Figure (1): Replacing a raised cosine channel with rectangular channel***

Therefore, the GN formula is given by [11]:

$$G_{NLI}(f) = \frac{16}{27} \sum_{m_{ch}=1}^{N_c} \sum_{n_{ch}=1}^{N_c} \sum_{k_{ch}=1}^{N_c} G_{m_{ch}} G_{n_{ch}} \int_{f_{s,n_{ch}}}^{f_{e,n_{ch}}} \int_{f_{s,m_{ch}}}^{f_{e,m_{ch}}} G_{k_{ch}}(f_1 + f_2 - f) \, |LK(f_1, f_2, f_1 + f_2 - f)|^2 \, df_1 df_2 \quad (1)$$

Where $N_c$ is the number of channels available in WDM comb and $LK(.)$ is the link function which is determined based on optical fiber link configuration. $f_{s,i}$ and $f_{e,i}$ are the start and end frequency of the $i$'th rectangular channel in WDM comb respectively. $G_n(f)$ is the PSD rectangular function due to $n$'th ($\forall n$) channel in the WDM comb which is launched to the first span of the fiber link while $G_n$ is the maximum value of $G_n(f)$ . Assuming $G_n(f)$ $\forall n$ has rectangular shape, (1) can be written as:

$$G_{NLI}(f) = \frac{16}{27} \sum_{m_{ch}=1}^{N_c} \sum_{n_{ch}=1}^{N_c} \sum_{k_{ch}=1}^{N_c} G_{m_{ch}} G_{n_{ch}} G_{k_{ch}} \times \iint_{S(m_{ch},n_{ch},k_{ch})} |LK(f_1, f_2, f_1 + f_2 - f)|^2 \, df_1 df_2 \quad (2)$$

Which in (2), $S(m_{ch}, n_{ch}, k_{ch})$ is the area confined by three criteria as below:

$$f_{s,m_{ch}} \le f_1 \le f_{e,m_{ch}} \quad (3)$$
$$f_{s,n_{ch}} \le f_2 \le f_{e,n_{ch}} \quad (4)$$
$$f_{s,k_{ch}} + f \triangleq f'_{s,k_{ch}} \le f_1 + f_2 \le f'_{e,k_{ch}} \triangleq f_{e,k_{ch}} + f \quad (5)$$

We call the area confined by (3)-(5) an integration island. In figure (2), we can see a typical scheme of the $S(m_{ch}, n_{ch}, k_{ch})$, hatched by blue color, in the $f_1 - f_2$ plane.

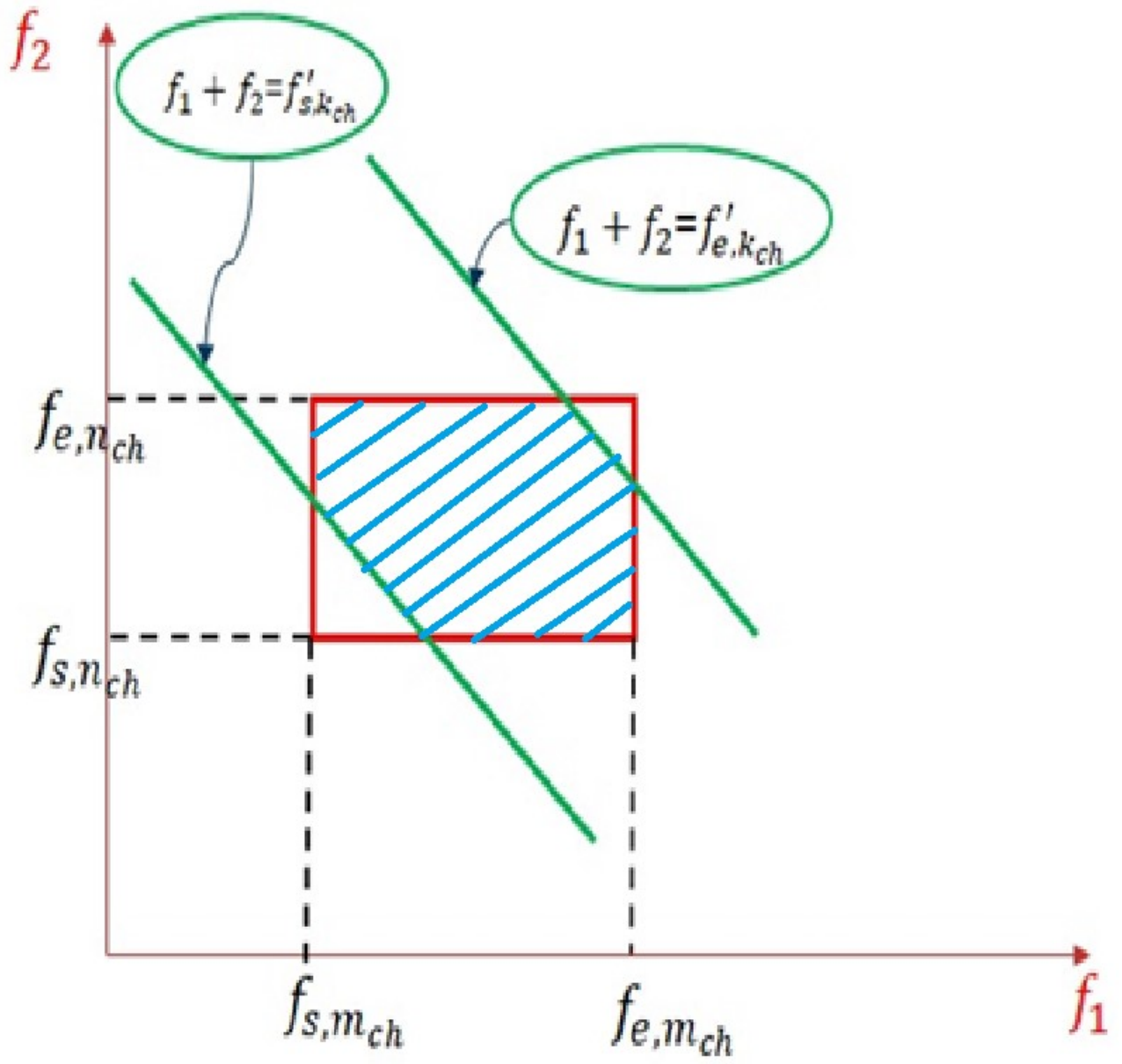

***Figure (2): The scheme of the formation of a typical Integration island in f1-f2 plane***

In general, the 2-D integral in (2) will be very complicated as different shapes may appear for each integration island. In this case, we proposed an approximated method in [11] which replaces the complex shape island with an square with same area and similar geometric center point as it is shown in figure (3) schematically.

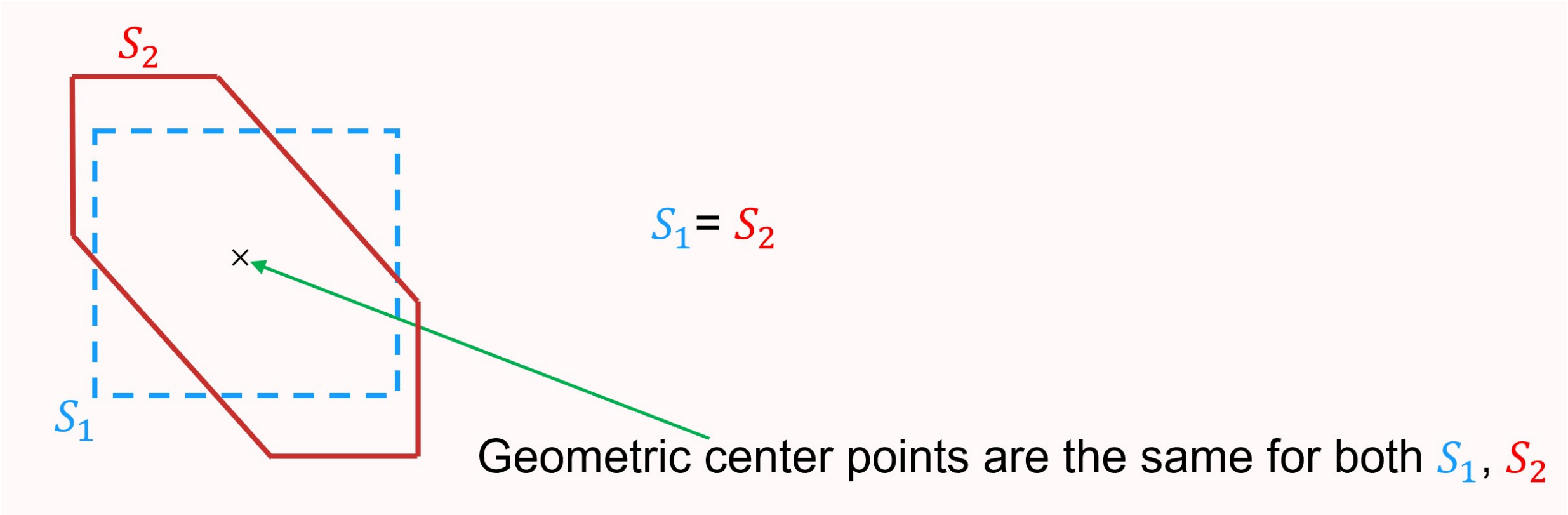

***Figure (3): replacing an arbitrary shape integration island with a concentric equivalent rectangle with the same area***

With this approximation, GN formula in equation (2) can be written as:

$$G_{NLI}(f) \cong \frac{16}{27} \sum_{m_{ch}=1}^{N_c} \sum_{n_{ch}=1}^{N_c} \sum_{k_{ch}=1}^{N_c} G_{m_{ch}} G_{n_{ch}} G_{k_{ch}} \times \int_{f_2^*-\frac{L_2}{2}}^{f_2^*+\frac{L_2}{2}} \int_{f_1^*-\frac{L_1}{2}}^{f_1^*+\frac{L_1}{2}} |LK(f_1, f_2, f_1 + f_2 - f)|^2 \, df_1 df_2 \tag{6}$$

In (6), $f_1^*$, $f_2^*$, $L_1$, $L_2$ are dependent to $m_{ch}$, $n_{ch}$ and $k_{ch}$ ($f_i^* = f_i^*(m_{ch}, n_{ch}, k_{ch})$ and $L_i = L_i(m_{ch}, n_{ch}, k_{ch})$ for i=1,2) and they are calculated by formulas derived in [11]. First, we define:

$$S_1(\tau) \triangleq \frac{\left(\tau - f_{s,m_{ch}} - f_{s,n_{ch}}\right)^2}{2} \tag{7}$$

$$f_1^{(1)}(\tau) \triangleq \frac{2f_{s,m_{ch}}}{3} + \frac{\tau}{3} - \frac{f_{s,n_{ch}}}{3} \tag{8}$$

$$f_2^{(1)}(\tau) \triangleq \frac{2f_{s,n_{ch}}}{3} + \frac{\tau}{3} - \frac{f_{s,m_{ch}}}{3} \tag{9}$$

$$BW_{m_{ch}} \triangleq f_{e,m_{ch}} - f_{s,m_{ch}} \tag{10}$$

$$BW_{n_{ch}} \triangleq f_{e,n_{ch}} - f_{s,n_{ch}} \tag{11}$$

$$u(x) = \begin{cases} 1 & x > 0 \\ \frac{1}{2} & x = 0 \\ 0 & x < 0 \end{cases} \tag{12}$$

$$S_2(\tau_1, \tau_2) \triangleq (\tau_2 - \tau_1) \times \min(BW_{m_{ch}}, BW_{n_{ch}}) \tag{13}$$

$$f_1^{(2)}(\tau_1,\tau_2) \triangleq \frac{f_{e,m_{ch}}+f_{s,m_{ch}}}{2} \times u(BW_{n_{ch}} - BW_{m_{ch}}) + \left(\frac{\tau_1+\tau_2}{2} - \frac{f_{e,n_{ch}}+f_{s,n_{ch}}}{2}\right) \times u(BW_{m_{ch}} - BW_{n_{ch}}) \quad (14)$$

$$f_2^{(2)}(\tau_1,\tau_2) \triangleq \frac{f_{e,n_{ch}}+f_{s,n_{ch}}}{2} \times u(BW_{m_{ch}} - BW_{n_{ch}}) + \left(\frac{\tau_1+\tau_2}{2} - \frac{f_{e,m_{ch}}+f_{s,m_{ch}}}{2}\right) \times u(BW_{n_{ch}} - BW_{m_{ch}}) \quad (15)$$

$$S_3(\tau) \triangleq \frac{\left(\tau - f_{e,m_{ch}} - f_{e,n_{ch}}\right)^2}{2} \quad (16)$$

$$f_1^{(3)}(\tau) \triangleq \frac{2f_{e,m_{ch}}}{3} + \frac{\tau}{3} - \frac{f_{e,n_{ch}}}{3} \quad (17)$$

$$f_2^{(3)}(\tau) \triangleq \frac{2f_{e,n_{ch}}}{3} + \frac{\tau}{3} - \frac{f_{e,m_{ch}}}{3} \quad (18)$$

$$f'_{s,k_{ch}} \triangleq f_{s,k_{ch}} + f \quad (19)$$

$$f'_{e,k_{ch}} \triangleq f_{e,k_{ch}} + f \quad (20)$$

$$F_1 \triangleq f_{s,m_{ch}} + f_{s,n_{ch}} \quad (21)$$

$$F_2 \triangleq \min\left(\left(f_{s,m_{ch}} + f_{e,n_{ch}}\right), \left(f_{e,m_{ch}} + f_{s,n_{ch}}\right)\right) \quad (22)$$

$$F_3 \triangleq \max\left(\left(f_{s,m_{ch}} + f_{e,n_{ch}}\right), \left(f_{e,m_{ch}} + f_{s,n_{ch}}\right)\right) \quad (23)$$

$$F_4 \triangleq f_{e,m_{ch}} + f_{e,n_{ch}} \quad (24)$$

$$\tau_1^+ \triangleq \min(f'_{e,k_{ch}}, F_2) \quad (25)$$

$$\tau_1^- \triangleq \max(f'_{s,k_{ch}}, F_1) \quad (26)$$

$$\tau_1 \triangleq \max(f'_{s,k_{ch}}, F_2) \quad (27)$$

$$\tau_2 \triangleq \min(f'_{e,k_{ch}}, F_3) \quad (28)$$

$$\tau_3^+ \triangleq \max(f'_{s,k_{ch}}, F_3) \quad (29)$$

$$\tau_3^- \triangleq \min(f'_{e,k_{ch}}, F_4) \quad (30)$$

$$S_1^+ \triangleq S_1(\tau_1^+) \times u\left(F_2 - f'_{s,k_{ch}}\right) \times u\left(f'_{e,k_{ch}} - F_1\right) \quad (31)$$

$$S_1^- \triangleq S_1(\tau_1^-) \times u\left(F_2 - f'_{s,k_{ch}}\right) \times u\left(f'_{e,k_{ch}} - F_1\right) \quad (32)$$

$$S_2 \triangleq S_2(\tau_1,\tau_2) \times u\left(F_3 - f'_{s,k_{ch}}\right) \times u\left(f'_{e,k_{ch}} - F_2\right) \quad (33)$$

$$S_3^+ \triangleq S_3(\tau_3^+) \times u\left(F_4 - f'_{s,k_{ch}}\right) \times u\left(f'_{e,k_{ch}} - F_3\right) \quad (34)$$

$$S_3^- \triangleq S_3(\tau_3^-) \times u\left(F_4 - f'_{s,k_{ch}}\right) \times u\left(f'_{e,k_{ch}} - F_3\right) \quad (35)$$

$$S = S_1^+ - S_1^- + S_2 + S_3^+ - S_3^- \quad (36)$$

Based on the above definitions, $f_1^*$, $f_2^*$, $L_1$, $L_2$ are given by [11]:

$$L_1 = L_2 = \sqrt{S} \quad (37)$$

$$f_1^* = \frac{S_1^+ \times f_1^{(1)}(\tau_1^+) - S_1^- \times f_1^{(1)}(\tau_1^-) + S_2 \times f_1^{(2)}(\tau_1,\tau_2) + S_3^+ \times f_1^{(3)}(\tau_3^+) - S_3^- \times f_1^{(3)}(\tau_3^-)}{S_1^+ - S_1^- + S_2 + S_3^+ - S_3^-} \quad (38)$$

$$f_2^* = \frac{S_1^+ \times f_2^{(1)}(\tau_1^+) - S_1^- \times f_2^{(1)}(\tau_1^-) + S_2 \times f_2^{(2)}(\tau_1, \tau_2) + S_3^+ \times f_2^{(3)}(\tau_3^+) - S_3^- \times f_2^{(3)}(\tau_3^-)}{S_1^+ - S_1^- + S_2 + S_3^+ - S_3^-} \quad (39)$$

It is worth noting that if $L_1 = 0$ or $L_2 = 0$ the 2-D integral in (6) is zero.

## 3. Fiber model

We assume that in the absence of the nonlinearity, the input electical field to a fiber span ($E_{in}(f)$) evolves in frequency domain during propagation as:

$$E(z,f) = E_{in}(f) \times e^{-j\int_0^z \beta_{n_s}(z',f)dz'} \times e^{-\int_0^z \alpha_{n_s}(z',f)dz'} \quad (40)$$

Where $\alpha_{n_s}(z,f)$ is the loss parameter for the $n_s'th$ fiber span which models three effects in the fiber i.e. ,propagation loss, possible distributed Raman amplification in the fiber and also stimulated Raman scattering (SRS) effect of the fiber. Also $\beta_{n_s}(z,f)$ in (40), is the propagation constant for the $n_s'th$ fiber span which handles all orders of dispersion. In this work we assume that the dispersion is only a function of frequency and not function of the distant so $\beta_{n_s}(z,f)$= $\beta_{n_s}(f)$. For the total loss we consider [9]:

$$\alpha_{n_s}(z,f) = \alpha_{0,n_s}(f)+\alpha_{1,n_s}(f) \times \exp(-\sigma_{n_s}(f) \times z) \quad (41)$$

Also for the propagation constant we only consider the Tylor series expansion in frequency domain up to order3 as:

$$\beta_{n_s}(z,f)= \beta_{n_s}(f) = \beta_{0,n_s} + 2\pi\beta_{1,n_s}\left(f - f_{n_s}^c\right) + 4\pi^2 \frac{\beta_{2,n_s}}{2}\left(f - f_{n_s}^c\right)^2 + 8\pi^3 \frac{\beta_{3,n_s}}{6}\left(f - f_{n_s}^c\right)^3 \quad (42)$$

In equation (42), $f_{n_s}^c$ is the center frequency for Tylor expansion and in general can be different span by span but during each span it must be constant. Also $\beta_{0,n_s}$ , $\beta_{1,n_s}$ , $\beta_{2,n_s}$ , $\beta_{3,n_s}$ are constant values along each span but they can change span by span.

Also at the end of each fiber span we consider an amplifier followed by a possible lumped accumulated dispersion element with input-output relation as:

$$E_{out}(f) = E_{in}(f)\left(\Gamma_{n_s}(f)\right)^{\frac{1}{2}} e^{j\theta_{n_s}(f)} e^{-j\beta_{DCU}^{(n_s)}(f)} \quad (43)$$

Where in (43), $\Gamma_{n_s}(f)$ is the power gain of the optical amplifier (EDFA) at the end of the $n_s'th$ fiber span and $\theta_{n_s}(f)$ is the phase imposed to the electrical field through the EDFA possible linear filtering property. $\beta_{DCU}^{(n_s)}(f)$ is the lumped accumulated dispersion at the end of the $n_s'th$ fiber span.

## 4. Incoherent approximation of the GN model

For analytical calculation of the GN formula in (6), we need to calculate $|LK(f_1, f_2, f_1 + f_2 - f)|^2$ which is calculated in equation (92) of [11] as:

$$|LK(f_1, f_2, f_3)|^2 \cong \sum_{n_s=1}^{N_s} \gamma_{n_s}^2 \times g_0^2(n_s) \times |\xi(n_s, f_1, f_2, f)|^2 +$$

$$\sum_{n_s=1}^{N_s} \sum_{n_s'=n_s+1}^{N_s} \gamma_{n_s}\gamma_{n_s'} \times g_0(n_s')g_0(n_s) \times 2 \times Real\{\xi(n_s, f_1, f_2, f)$$

$$\times \xi^*(n_s', f_1, f_2, f) \times e^{j[4\pi^2(f_1-f)(f_2-f)\times\Delta\beta_{acc}(n_s,n_s',f_1,f_2)+\Delta\theta(n_s,n_s')]}\}$$

The incoherent approximation of the GN model ignores the interaction of the spans in the integrand function of the GN 2-D integral and therefore the second part of right side of the above equation is disregarded and $|LK(f_1, f_2, f_3)|^2$ is approximated as:

$$|LK(f_1, f_2, f_3)|^2 \cong \sum_{n_s=1}^{N_s} \gamma_{n_s}^2 \times g_0^2(n_s, f_1^*, f_2^*, f) \times |\xi(n_s, f_1, f_2, f)|^2 \quad (44)$$

Where in (44) . $\gamma_{n_s}$is the nonlinearity parameter of the $n_s'th$ fiber span and $g_0(n_s)$ and $\xi(n_s, f_1, f_2, f)$ are defined as [11]:

$$g_0(n_s, f_1^*, f_2^*, f) \triangleq \prod_{p=1}^{n_s-1} \left\{ \left[\Gamma_p(f_1^*)\Gamma_p(f_2^*)\Gamma_p(f_3^*)\right]^{\frac{1}{2}} \times e^{-L_s(p)\sum_{i=1}^{3}\alpha_{0,p}(f_i^*)} \times e^{\sum_{i=1}^{3}\left\{\frac{\alpha_{1,p}(f_i^*)}{\sigma_p(f_i^*)}\exp\left(-\sigma_p(f_i^*)\times L_s(p)\right) - \frac{\alpha_{1,p}(f_i^*)}{\sigma_p(f_i^*)}\right\}} \right\} \times \quad (45)$$

$$\prod_{p=n_s}^{N_s} \left\{ \left(\Gamma_p(f)\right)^{\frac{1}{2}} \times e^{-\left(\alpha_{0,p}(f)L_s(p)+\frac{\alpha_{1,p}(f)}{\sigma_p(f)} - \frac{\alpha_{1,p}(f)}{\sigma_p(f)}\exp\left(-\sigma_p(f)\times L_s(p)\right)\right)} \right\}$$

$$\xi(n_s, f_1, f_2, f) \triangleq \left\{ \frac{1}{2\overline{\alpha_{0,n_s}} - j4\pi^2(f_1 - f)(f_2 - f)\left(\beta_{2,n_s} + \pi\beta_{3,n_s}(f_1 + f_2 - 2f_{n_s}^c)\right)} - \frac{2\overline{\alpha_{1,n_s}}}{\left[2\overline{\alpha_{0,n_s}} - j4\pi^2(f_1 - f)(f_2 - f)\left(\beta_{2,n_s} + \pi\beta_{3,n_s}(f_1 + f_2 - 2f_{n_s}^c)\right)\right]} \times \frac{1}{\left[2\overline{\alpha_{0,n_s}} + \overline{\sigma_{n_s}} - j4\pi^2(f_1 - f)(f_2 - f)\left(\beta_{2,n_s} + \pi\beta_{3,n_s}(f_1 + f_2 - 2f_{n_s}^c)\right)\right]} \right\} \tag{46}$$

It is worth noticing that $G_{m_{ch}}$ is the maximum value of the PSD function entering the first span. But it will be scaled at the entrance of the $n_s$'th span as:

$$G_{m_{ch}}^{(n_s)} \cong G_{m_{ch}} \times \prod_{p=1}^{n_s-1} \left\{ \Gamma_p(f_1^*) \times e^{-2L_s(p)\alpha_{0,p}(f_1^*)} \times e^{\frac{2\alpha_{1,p}(f_1^*)}{\sigma_p(f_1^*)}\exp\left(-\sigma_p(f_1^*)\times L_s(p)\right) - \frac{2\alpha_{1,p}(f_1^*)}{\sigma_p(f_1^*)}} \right\}$$

$$G_{n_{ch}}^{(n_s)} \cong G_{n_{ch}} \times \prod_{p=1}^{n_s-1} \left\{ \Gamma_p(f_2^*) \times e^{-2L_s(p)\alpha_{0,p}(f_2^*)} \times e^{\frac{2\alpha_{1,p}(f_2^*)}{\sigma_p(f_2^*)}\exp\left(-\sigma_p(f_2^*)\times L_s(p)\right) - \frac{2\alpha_{1,p}(f_2^*)}{\sigma_p(f_2^*)}} \right\}$$

$$G_{k_{ch}}^{(n_s)} \cong G_{k_{ch}} \times \prod_{p=1}^{n_s-1} \left\{ \Gamma_p(f_3^*) \times e^{-2L_s(p)\alpha_{0,p}(f_3^*)} \times e^{\frac{2\alpha_{1,p}(f_3^*)}{\sigma_p(f_3^*)}\exp\left(-\sigma_p(f_3^*)\times L_s(p)\right) - \frac{2\alpha_{1,p}(f_3^*)}{\sigma_p(f_3^*)}} \right\}$$

Where in the above equation and (45) $f_3^* = f_1^* + f_2^* - f$ and $L_s(p)$ is the length of the $p'th$ span in the fiber link. Also $\overline{\alpha_{0,n_s}} = \overline{\alpha_{0,n_s}}(f_1^*, f_2^*, f)$ in (46) is given by [11]:

$$\overline{\alpha_{0,n_s}} = \frac{\alpha_{0,n_s}(f_1^*) + \alpha_{0,n_s}(f_2^*) + \alpha_{0,n_s}(f_3^*) - \alpha_{0,n_s}(f)}{2} \tag{47}$$

Furthermore, $\overline{\alpha_{1,n_s}} = \overline{\alpha_{1,n_s}}(f_1^*, f_2^*, f)$ and $\overline{\sigma_{n_s}}=\overline{\sigma_{n_s}}(f_1^*, f_2^*, f)$ in (46) are found based on the basic assumption [11] that equation (48) holds for $0 \leq z \leq L_s(n_s)$ as:

$$\overline{\alpha_{1,n_s}} \exp(-\overline{\sigma_{n_s}} z) \cong \frac{1}{2} \times \{\alpha_{1,n_s}(f_1^*)\exp(-\sigma_{n_s}(f_1^*)\, z) + \alpha_{1,n_s}(f_2^*)\exp(-\sigma_{n_s}(f_2^*)\, z) - \alpha_{1,n_s}(f)\exp(-\sigma_{n_s}(f)\, z) + \alpha_{1,n_s}(f_3^*)\exp(-\sigma_{n_s}(f_3^*)\, z)\} \quad \text{for } 0 \leq z \leq L_s(n_s) \tag{48}$$

To continue the derivation of closed form formula, we accept an approximation for $\xi(n_s, f_1, f_2, f)$ in equation (46) as:

$$\xi(n_s, f_1, f_2, f) \cong \left\{ \frac{1}{2\overline{\alpha_{0,n_s}} - j4\pi^2(f_1 - f)(f_2 - f)\overline{\overline{\beta_{2,n_s}}}} - \frac{2\overline{\alpha_{1,n_s}}}{\left[2\overline{\alpha_{0,n_s}} - j4\pi^2(f_1 - f)(f_2 - f)\overline{\beta_{2,n_s}}\right]} \times \frac{1}{\left[2\overline{\alpha_{0,n_s}} + \overline{\sigma_{n_s}} - j4\pi^2(f_1 - f)(f_2 - f)\overline{\beta_{2,n_s}}\right]} \right\} \quad (49)$$

Where $\overline{\beta_{2,n_s}}$ in (49) is independent of and is given by [9],[10]:

$$\overline{\beta_{2,n_s}} \triangleq \left(\beta_{2,n_s} + \pi\beta_{3,n_s}\left(f_1^* + f_2^* - 2f_{n_s}^c\right)\right) \quad (50)$$

Also, $|\xi(n_s, f_1, f_2, f)|^2$ can be calculated using (49) as:

$$|\xi(n_s, f_1, f_2, f)|^2 \cong \frac{J_1(n_s)}{1+(f_1-f)^2(f_2-f)^2\left(\overline{D_1}(n_s)\right)^2} + \frac{J_2(n_s)}{1+(f_1-f)^2(f_2-f)^2\left(\overline{D_2}(n_s)\right)^2} \quad (51)$$

Where $J_1(n_s)$, $J_2(n_s)$, $\overline{D_1}(n_s)$ and $\overline{D_2}(n_s)$ in (51) are given by [11]:

$$J_1(n_s, f_1^*, f_2^*, f) = \frac{4\overline{\alpha_{1,n_s}} \times (2\overline{\alpha_{0,n_s}} - \overline{\alpha_{1,n_s}} + \overline{\sigma_{n_s}})}{\overline{\sigma_{n_s}} \times \left(2\overline{\alpha_{0,n_s}} + \overline{\sigma_{n_s}}\right)^2 \times (4\overline{\alpha_{0,n_s}} + \overline{\sigma_{n_s}})} \quad (52)$$

$$J_2(n_s, f_1^*, f_2^*, f) = \frac{(\overline{\sigma_{n_s}} - 2\overline{\alpha_{1,n_s}}) \times (4\overline{\alpha_{0,n_s}} - 2\overline{\alpha_{1,n_s}} + \overline{\sigma_{n_s}})}{4\overline{\sigma_{n_s}} \times \left(\overline{\alpha_{0,n_s}}\right)^2 \times (4\overline{\alpha_{0,n_s}} + \overline{\sigma_{n_s}})} \quad (53)$$

$$\overline{D_1}(n_s, f_1^*, f_2^*, f) = \frac{4\pi^2\overline{\beta_{2,n_s}}}{2\overline{\alpha_{0,n_s}} + \overline{\sigma_{n_s}}} \quad (54)$$

$$\overline{D_2}(n_s, f_1^*, f_2^*, f) = \frac{2\pi^2\overline{\beta_{2,n_s}}}{\overline{\alpha_{0,n_s}}} \quad (55)$$

Using (51), we can rewrite the noncoherent approximation of the link function presented in equation (44) as follows:

$$|LK(f_1,f_2,f_3)|^2 \cong \sum_{n_s=1}^{N_s} \gamma_{n_s}^2 \times g_0^2(n_s,f_1^*,f_2^*,f) \times \left\{\frac{J_1(n_s,f_1^*,f_2^*,f)}{1+(f_1-f)^2(f_2-f)^2\left(\overline{D_1}(n_s,f_1^*,f_2^*,f)\right)^2} + \frac{J_2(n_s,f_1^*,f_2^*,f)}{1+(f_1-f)^2(f_2-f)^2\left(\overline{D_2}(n_s,f_1^*,f_2^*,f)\right)^2}\right\} \tag{56}$$

Therefore, with applying (56), equation (6) can be written as:

$$G_{NLI}(f) \cong \frac{16}{27}\sum_{m_{ch}=1}^{N_c}\sum_{n_{ch}=1}^{N_c}\sum_{k_{ch}=1}^{N_c} G_{m_{ch}}G_{n_{ch}}G_{k_{ch}} \times \sum_{n_s=1}^{N_s}\gamma_{n_s}^2 \times g_0^2(n_s,f_1^*,f_2^*,f) \times \int_{f_2^*-\frac{L_2}{2}}^{f_2^*+\frac{L_2}{2}}\int_{f_1^*-\frac{L_1}{2}}^{f_1^*+\frac{L_1}{2}} \left\{\frac{J_1(n_s,f_1^*,f_2^*,f)}{1+(f_1-f)^2(f_2-f)^2\left(\overline{D_1}(n_s,f_1^*,f_2^*,f)\right)^2} + \frac{J_2(n_s,f_1^*,f_2^*,f)}{1+(f_1-f)^2(f_2-f)^2\left(\overline{D_2}(n_s,f_1^*,f_2^*,f)\right)^2}\right\} df_1 df_2 \tag{57}$$

Changing the integration variables in (57) with $f_1' = f_1 - f$ and $f_2' = f_2 - f$ we have:

$$G_{NLI}(f) \cong \frac{16}{27}\sum_{m_{ch}=1}^{N_c}\sum_{n_{ch}=1}^{N_c}\sum_{k_{ch}=1}^{N_c} G_{m_{ch}}G_{n_{ch}}G_{k_{ch}} \times \sum_{n_s=1}^{N_s}\gamma_{n_s}^2 \times g_0^2(n_s,f_1^*,f_2^*,f) \times \int_{f_2^*-f-\frac{L_2}{2}}^{f_2^*-f+\frac{L_2}{2}}\int_{f_1^*-f-\frac{L_1}{2}}^{f_1^*-f+\frac{L_1}{2}} \left\{\frac{J_1(n_s,f_1^*,f_2^*,f)}{1+f_1'^2\, f_2'^2\left(\overline{D_1}(n_s,f_1^*,f_2^*,f)\right)^2} + \frac{J_2(n_s,f_1^*,f_2^*,f)}{1+f_1'^2\, f_2'^2\left(\overline{D_2}(n_s,f_1^*,f_2^*,f)\right)^2}\right\} df_1' df_2' \tag{58}$$

For reaching a closed form formula, the 2-D integral in (58) must be solved analytically. We can see that if $A, B, X_1, X_2, Y_1, Y_2$ are constant values:

$$\int_{Y_1}^{Y_2}\int_{X_1}^{X_2} \frac{B}{1+x^2\, y^2 A^2}\, dxdy = \frac{B}{2A}\left(F_{int}(AX_1Y_1) + F_{int}(AX_2Y_2) - F_{int}(AX_2Y_1) - F_{int}(AX_1Y_2)\right) \tag{59}$$

Where $F_1(.)$ In (59) is:

$$F_{int}(x) \triangleq j \times \{Li_2(-jx) - Li_2(jx)\} \tag{60}$$

$Li_2(.)$ Is the second order polylogarithm function. Therefore, (58) can be written as:

$$G_{NLI}(f) \cong \frac{16}{27} \sum_{m_{ch}=1}^{N_c} \sum_{n_{ch}=1}^{N_c} \sum_{k_{ch}=1}^{N_c} G_{m_{ch}} G_{n_{ch}} G_{k_{ch}} \times \sum_{n_s=1}^{N_s} \gamma_{n_s}^2 \times g_0^2(n_s, f_1^*, f_2^*, f) \times \sum_{i=1}^{2} \frac{J_i(n_s, f_1^*, f_2^*, f)}{2\overline{D}_\iota(n_s, f_1^*, f_2^*, f)} \times \sum_{j=1}^{4} (-1)^{(\delta_{j,3}+\delta_{j,4})} F_{int}\left(\overline{D}_\iota(n_s, f_1^*, f_2^*, f) \times \left(f_2^* - f + (-1)^j\, \frac{L_2}{2}\right) \times \left(f_1^* - f + (-1)^{(\delta_{j,1}+\delta_{j,4})}\, \frac{L_1}{2}\right)\right) \tag{61}$$

Where $\delta_{m,n}$ in (61) is the Kronecker delta function and is defined as:

$$\delta_{m,n} = \begin{cases} 1 & for\ m = n \\ 0 & for\ m \neq n \end{cases} \tag{62}$$

The channel under test (CUT) is the channel that $f$ is located in it. The PSD in the center frequency of the CUT ($f_{CUT} = \frac{f_{s,CUT}+f_{e,CUT}}{2}$) can be calculated by replacing $f$ with $f_{CUT}$ in equation (61) as:

$$G_{NLI}(f_{CUT}) \cong \frac{16}{27} \sum_{m_{ch}=1}^{N_c} \sum_{n_{ch}=1}^{N_c} \sum_{k_{ch}=1}^{N_c} G_{m_{ch}} G_{n_{ch}} G_{k_{ch}} \times \sum_{n_s=1}^{N_s} \gamma_{n_s}^2 \times g_0^2(n_s, f_1^*, f_2^*, f_{CUT}) \times \sum_{i=1}^{2} \frac{J_i(n_s, f_1^*, f_2^*, f_{CUT})}{2\overline{D}_\iota(n_s, f_1^*, f_2^*, f_{CUT})} \times \sum_{j=1}^{4} (-1)^{(\delta_{j,3}+\delta_{j,4})} F_{int}\left(\overline{D}_\iota(n_s, f_1^*, f_2^*, f_{CUT}) \times \left(f_2^* - f_{CUT} + (-1)^j \frac{L_2}{2}\right) \times \left(f_1^* - f_{CUT} + (-1)^{(\delta_{j,1}+\delta_{j,4})} \frac{L_1}{2}\right)\right) \tag{63}$$

$$S = \{(m_{ch}, n_{ch}, k_{ch}) \mid \quad 1 \le m_{ch} \le N_c\ ,\ 1 \le n_{ch} \le N_c\ , 1 \le k_{ch} \le N_c\}$$

$$S^{SCI} = \{(m_{ch}, n_{ch}, k_{ch}) \in S \mid m_{ch} = n_{ch} = k_{ch} = CUT\}$$

$$S^{XCI,1} = \{(m_{ch}, n_{ch}, k_{ch}) \in S \mid m_{ch} = n_{ch} = CUT, k_{ch} \ne CUT\}$$

$$S^{XCI,2} = \{(m_{ch}, n_{ch}, k_{ch}) \in S \mid m_{ch} = k_{ch} = CUT, n_{ch} \ne CUT\}$$

$$S^{XCI,3} = \{(m_{ch}, n_{ch}, k_{ch}) \in S \mid k_{ch} = n_{ch} = CUT, m_{ch} \ne CUT\}$$

$$S^{XCI,4} = \{(m_{ch}, n_{ch}, k_{ch}) \in S \mid k_{ch} = n_{ch} \ne CUT, m_{ch} = CUT\}$$

$$S^{XCI,5} = \{(m_{ch}, n_{ch}, k_{ch}) \in S \mid k_{ch} = m_{ch} \ne CUT, n_{ch} = CUT\}$$

$$S^{XCI,6} = \{(m_{ch}, n_{ch}, k_{ch}) \in S \mid m_{ch} = n_{ch} \ne CUT, k_{ch} = CUT\}$$

$$\begin{aligned} S^{XCI} &= S^{XCI,1} \cup S^{XCI,2} \cup S^{XCI,3} \cup S^{XCI,4} \cup S^{XCI,5} \cup S^{XCI,6} \\ &= \{(m_{ch}, n_{ch}, k_{ch}) \in S \mid (m_{ch}, n_{ch}, k_{ch}) \in S^{XCI,1}\ or\ (m_{ch}, n_{ch}, k_{ch}) \\ &\in S^{XCI,2}\ or\ (m_{ch}, n_{ch}, k_{ch}) \in S^{XCI,3}\ or\ (m_{ch}, n_{ch}, k_{ch}) \in S^{XCI,4} or\ (m_{ch}, n_{ch}, k_{ch}) \\ &\in S^{XCI,5}\ or\ (m_{ch}, n_{ch}, k_{ch}) \in S^{XCI,6}\} \end{aligned}$$

$$\begin{aligned} S^{SCI-XCI} &= S^{SCI} \cup S^{XCI} \\ &= \{(m_{ch}, n_{ch}, k_{ch}) \in S \mid (m_{ch}, n_{ch}, k_{ch}) \in S^{SCI}\ or\ (m_{ch}, n_{ch}, k_{ch}) \in S^{XCI}\} \end{aligned}$$

$$S^{MCI} = S - S^{SCI-XCI} = S \cap \overline{S^{SCI-XCI}} = \{(m_{ch}, n_{ch}, k_{ch}) \in S \mid (m_{ch}, n_{ch}, k_{ch}) \notin S^{SCI-XCI}\}$$

$$S = S^{MCI} \cup S^{SCI-X} \quad , S^{MCI} \cap S^{SCI-XCI} = \emptyset$$

The NLI in (63) has three contributions:

(1)- The self-channel interference (SCI) that is equivalent to $m_{ch} = n_{ch} = k_{ch} = CUT$. ($S^{SCI}$)

(2)-The cross-channel interference (XCI) that is equivalent to two terms: ($S^{XCI}$)

(2-a) one variable among three ( $m_{ch}, n_{ch}, k_{ch}$) is equal to $CUT$ while two others are equal to each other but not equal to CUT.( $S^{XCI,4} \cup S^{XCI,5} \cup S^{XCI,6}$)

(2-b) two variables among three ( $m_{ch}, n_{ch}, k_{ch}$) are equal to $CUT$ while the other one is not equal to CUT.( $S^{XCI,1} \cup S^{XCI,2} \cup S^{XCI,3}$)

(3)- The multi-channel interference (MCI) terms that are all possible ( $m_{ch}, n_{ch}, k_{ch}$) which are not in SCI and XCI terms.( $S^{MCI}$)

The SCI and XCI contribution of NLI can be approximately represented as derived in [3] by setting:

$$S^{SC} \quad \cong \{(m_{ch}, n_{ch}, k_{ch}) \in S \mid \{\ m_{ch} = n_{ch} = k_{ch} = CUT\}$$
$$or\ \{m_{ch} = CUT\ \& n_{ch} = k_{ch} \neq CUT\}\ or\ \{n_{ch} = CUT\ \& m_{ch} = k_{ch} \neq CUT\}\}$$

$f_2^* = f_{CUT}, L_2 = BW_{CUT}, f_1^* = f_3^* = \frac{f_{s,m_{ch}} + f_{e,m_{ch}}}{2}, L_1 = BW_{m_{ch}}$.

For finding $\overline{\alpha_{0,n_s}}$, $\overline{\alpha_{1,n_s}}$, $\overline{\sigma_{n_s}}$ and $\overline{\beta_{2,n_s}}$ for SCI-XCI contributions, we set $f_2^* = f_{CUT}$, $f_1^* = f_3^* = \frac{f_{s,m_{ch}} + f_{e,m_{ch}}}{2}$ in (47), (48) and (50). Therefore we will have :

$$\overline{\alpha_{0,n_s}} = \alpha_{0,n_s}(\frac{f_{s,m_{ch}} + f_{e,m_{ch}}}{2}) \tag{64}$$

$$\overline{\alpha_{1,n_s}} = \alpha_{1,n_s}\left(\frac{f_{s,m_{ch}} + f_{e,m_{ch}}}{2}\right) \tag{65}$$

$$\overline{\sigma_{n_s}} = \sigma_{n_s}\left(\frac{f_{s,m_{ch}} + f_{e,m_{ch}}}{2}\right) \tag{66}$$

$$\overline{\beta_{2,n_s}} \triangleq \left(\beta_{2,n_s} + \pi\beta_{3,n_s}\left(\frac{f_{s,m_{ch}} + f_{e,m_{ch}}}{2} + f_{CUT} - 2f_{n_s}^c\right)\right) \tag{67}$$

The SCI and XCI contribution of NLI is denoted by $G_{NLI}^{SCI-X}\ (f_{CUT})$ and similar to formula (41) in [3] can be written as:

$$G_{NLI}^{SCI-XCI}(f_{CUT}) \cong \frac{16}{27}\sum_{m_{ch}=1}^{N_c} G_{m_{ch}}^2 G_{CUT}\left(2-\delta_{m_{ch},CUT}\right)\times \tag{68}$$

$$\sum_{n_s=1}^{N_s}\gamma_{n_s}^2\times g_0^2\left(n_s,\frac{f_{s,m_{ch}}+f_{e,m_{ch}}}{2},f_{CUT},f_{CUT}\right)$$

$$\times\sum_{i=1}^{2}\frac{J_i(n_s,\frac{f_{s,m_{ch}}+f_{e,m_{ch}}}{2},f_{CUT},f_{CUT})}{\overline{D_\iota}\left(n_s,\frac{f_{s,m_{ch}}+f_{e,m_{ch}}}{2},f_{CUT},f_{CUT}\right)}$$

$$\times\sum_{j=1}^{2}(-1)^j F_{int}\left(\overline{D_\iota}\left(n_s,\frac{f_{s,m_{ch}}+f_{e,m_{ch}}}{2},f_{CUT},f_{CUT}\right)\times\frac{BW_{CUT}}{2}\right.$$

$$\left.\times\left(\frac{f_{s,m_{ch}}+f_{e,m_{ch}}}{2}-f_{CUT}+(-1)^j\frac{BW_{m_{ch}}}{2}\right)\right)$$

For calculation of $G_{NLI}^{SCI-XCI}(f_{CUT})$ from (68), we first calculate $\overline{\alpha_{0,n_s}}$, $\overline{\alpha_{1,n_s}}$, $\overline{\sigma_{n_s}}$ and $\overline{\beta_{2,n_s}}$ by (64)-(67) and then $J_i(n_s,\frac{f_{s,m_{ch}}+f_{e,m_{ch}}}{2},f_{CUT},f_{CUT})$ and $\overline{D_\iota}\left(n_s,\frac{f_{s,m_{ch}}+f_{e,m_{ch}}}{2},f_{CUT},f_{CUT}\right)$ can be calculated through (52)-(55). Also for calculating $g_0\left(n_s,\frac{f_{s,m_{ch}}+f_{e,m_{ch}}}{2},f_{CUT},f_{CUT}\right)$ in (68) we replace $f_2^*=f=f_{CUT}$ and $f_1^*=f_3^*=\frac{f_{s,m_{ch}}+f_{e,m_{ch}}}{2}$ in (45).

The MCI contribution of NLI is denoted by $G_{NLI}^{MCI}(f_{CUT})$ . It is like equation (63) but the summations are hold in the MCI terms as:

$$G_{NLI}^{MCI}(f_{CUT}) \cong \frac{16}{27} \sum_{m_{ch}=1}^{N_c} \sum_{n_{ch}=1}^{N_c} \sum_{k_{ch}=1}^{N_c} G_{m_{ch}} G_{n_{ch}} G_{k_{ch}} \times \quad (m_{ch}, n_{ch}, k_{ch}) \in S^{MCI} \sum_{n_s=1}^{N_s} \gamma_{n_s}^2 \times g_0^2(n_s, f_1^*, f_2^*, f_{CUT}) \times \sum_{i=1}^{2} \frac{J_i(n_s, f_1^*, f_2^*, f_{CUT})}{2\overline{D_i}(n_s, f_1^*, f_2^*, f_{CUT})} \times \sum_{j=1}^{4} (-1)^{(\delta_{j,3}+\delta_{j,4})} F_{int}\left(\overline{D_i}(n_s, f_1^*, f_2^*, f_{CUT}) \times \left(f_2^* - f_{CUT} + (-1)^j \frac{L_2}{2}\right) \times \left(f_1^* - f_{CUT} + (-1)^{(\delta_{j,1}+\delta_{j,4})} \frac{L_1}{2}\right)\right) \quad (69)$$

Where in (69), $f_1^* = f_1^*(m_{ch}, n_{ch}, k_{ch})$, $f_2^* = f_2^*(m_{ch}, n_{ch}, k_{ch})$, $L_1 = L_1(m_{ch}, n_{ch}, k_{ch})$ and $L_2 = L_2(m_{ch}, n_{ch}, k_{ch})$ and they are calculated using (7)-(39) replacing $f = f_{CUT}$ . Also $g_0(n_s, f_1^*, f_2^*, f_{CUT})$ in (69) is calculated by replacing $f = f_{CUT}$ in (45). For calculating $J_i(n_s, f_1^*, f_2^*, f_{CUT})$ and $\overline{D_i}(n_s, f_1^*, f_2^*, f_{CUT})$ in (69), we first replace $f = f_{CUT}$ in (47), (48) and (50) to obtain $\overline{\alpha_{0,n_s}}$, $\overline{\alpha_{1,n_s}}$, $\overline{\sigma_{n_s}}$ and $\overline{\beta_{2,n_s}}$. Then, we use (52)-(55) for calculating $J_i(n_s, f_1^*, f_2^*, f_{CUT})$ and $\overline{D_i}(n_s, f_1^*, f_2^*, f_{CUT})$.

The total NLI is the sum of (68) and (69) as:

$$G_{NLI}(f_{CUT}) \cong G_{NLI}^{SCI-XCI}(f_{CUT}) + G_{NLI}^{MCI}(f_{CUT}) \quad (70)$$

## 5. Frequency and distance independent loss

If we consider the fiber attenuation model in the simplest form as:

$$\alpha_{n_s}(z, f) = \alpha_{0,n_s} \quad (71)$$

Where $\alpha_{0,n_s}$ is a scaler value independent of z and f but it depends to $n_s$. Therefore, we can rewrite (47) as:

$$\overline{\alpha_{0,n_s}} = \alpha_{0,n_s} \quad (72)$$

Also based on (48), we have:

$$\overline{\alpha_{1,n_s}} = 0 \tag{73}$$

Therefore, replacing (72) and (73) in (52)-(55) we will have:

$$J_1\left(n_s, f_1^*, f_2^*, f\right) = 0 \tag{74}$$

$$J_2\left(n_s, f_1^*, f_2^*, f\right) = \frac{1}{4 \times \left(\alpha_{0,n_s}\right)^2} \tag{75}$$

$$\overline{D_1}\left(n_s, f_1^*, f_2^*, f\right) = \frac{4\pi^2 \overline{\beta_{2,n_s}}}{2\alpha_{0,n_s} + \overline{\sigma_{n_s}}} \tag{76}$$

$$\overline{D_2}\left(n_s, f_1^*, f_2^*, f\right) = \frac{2\pi^2 \overline{\beta_{2,n_s}}}{\alpha_{0,n_s}} \tag{77}$$

Furthermore, by considering (72), (45) can be obtained as:

$$g_0(n_s, f_1^*, f_2^*, f) \triangleq \prod_{p=1}^{n_s-1} \left\{ \left[ \Gamma_p(f_1^*) \Gamma_p(f_2^*) \Gamma_p(f_3^*) \right]^{\frac{1}{2}} \times \ e^{-3\alpha_{0,n_s} L_s(p)} \right\} \times \prod_{p=n_s}^{N_s} \left\{ \left( \Gamma_p(f) \right)^{\frac{1}{2}} \times e^{-\alpha_{0,n_s} L_s(p)} \right\} \tag{78}$$

Therefore, (68) can be written as:

$$G_{NLI}^{SCI-XCI}(f_{CUT}) \cong \frac{16}{27}\sum_{m_{ch}=1}^{N_c} G_{m_{ch}}^2 G_{CUT}\left(2-\delta_{m_{ch},CUT}\right)\times \tag{79}$$

$$\sum_{n_s=1}^{N_s}\gamma_{n_s}^2\times g_0^2\left(n_s,\frac{f_{s,m_{ch}}+f_{e,m_{ch}}}{2},f_{CUT},f_{CUT}\right)\times\frac{1}{8\pi^2\alpha_{0,n_s}\overline{\beta_{2,n_s}}}$$

$$\times\sum_{j=1}^{2}(-1)^j F_{int}\left(\frac{\pi^2\overline{\beta_{2,n_s}}}{\alpha_{0,n_s}}\times BW_{CUT}\right.$$

$$\left.\times\left(\frac{f_{s,m_{ch}}+f_{e,m_{ch}}}{2}-f_{CUT}+(-1)^j\frac{BW_{m_{ch}}}{2}\right)\right)$$

For evaluation of $G_{NLI}^{SCI-XCI}(f_{CUT})$ in (79), $g_0\left(n_s,\frac{f_{s,m_{ch}}+f_{e,m_{ch}}}{2},f_{CUT},f_{CUT}\right)$ is calculated using (78) with replacing $f_1^*=f_3^*=\frac{f_{s,m_{ch}}+f_{e,m_{ch}}}{2}$ and $f_2^*=f=f_{CUT}$ in (78). Also, $\overline{\beta_{2,n_s}}$ in (79) is the same as equation (67) as:

$$\overline{\beta_{2,n_s}} \triangleq \left(\beta_{2,n_s}+\pi\beta_{3,n_s}\left(\frac{f_{s,m_{ch}}+f_{e,m_{ch}}}{2}+f_{CUT}-2f_{n_s}^c\right)\right) \tag{80}$$

Also, MCI contribution presented in (69), assuming (71), can be written as:

$$G_{NLI}^{MCI}(f_{CUT}) \cong \frac{16}{27} \sum_{\substack{m_{ch}=1 \\ (m_{ch}, n_{ch}, k_{ch}) \in S^{MCI}}}^{N_c} \sum_{n_{ch}=1}^{N_c} \sum_{k_{ch}=1}^{N_c} G_{m_{ch}} G_{n_{ch}} G_{k_{ch}} \times \sum_{n_s=1}^{N_s} \gamma_{n_s}^2 \times g_0^2(n_s, f_1^*, f_2^*, f_{CUT}) \times \frac{1}{16\pi^2 \alpha_{0,n_s} \overline{\beta_{2,n_s}}} \times \sum_{j=1}^{4} (-1)^{(\delta_{j,3}+\delta_{j,4})} F_{int}\left( \frac{2\pi^2 \overline{\beta_{2,n_s}}}{\alpha_{0,n_s}} \times \left(f_2^* - f_{CUT} + (-1)^j \frac{L_2}{2}\right) \times \left(f_1^* - f_{CUT} + (-1)^{(\delta_{j,1}+\delta_{j,4})} \frac{L_1}{2}\right)\right) \tag{81}$$

Where in (81), $f_1^* = f_1^*(m_{ch}, n_{ch}, k_{ch})$, $f_2^* = f_2^*(m_{ch}, n_{ch}, k_{ch})$, $L_1 = L_1(m_{ch}, n_{ch}, k_{ch})$ and $L_2 = L_2(m_{ch}, n_{ch}, k_{ch})$ and they are calculated using (7)-(39) replacing $f = f_{CUT}$ . Also $g_0(n_s, f_1^*, f_2^*, f_{CUT})$ in (69) is calculated by replacing $f = f_{CUT}$ in (78). Also, $\overline{\beta_{2,n_s}}$ in (81) is the same as (50):

$$\overline{\beta_{2,n_s}} \triangleq \left(\beta_{2,n_s} + \pi\beta_{3,n_s}\left(f_1^* + f_2^* - 2f_{n_s}^c\right)\right) \tag{82}$$

## 6. Introducing correction factors to $G_{NLI}^{SCI-XCI}(f_{CUT})$

In [10],[12], some corrections are imposed on (79). To introducing these corrections, we first rewrite (79) as:

$$G_{NLI}^{SCI-XCI}(f_{CUT}) \cong \tag{83}$$

$$\frac{16}{27}G_{CUT}^{3}\sum_{n_s=1}^{N_s}\gamma_{n_s}^{2}\times g_0^2(n_s,f_{CUT},f_{CUT},f_{CUT})\times\frac{F_{int}\left(\frac{\pi^2\overline{\beta_{2,n_s,SCI}}}{2\alpha_{0,n_s}}\times BW_{CUT}^2\right)}{4\pi^2\alpha_{0,n_s}\overline{\beta_{2,n_s,SCI}}}$$

$$+\frac{16}{27}\sum_{\substack{m_{ch}=1\\ m_{ch}\neq CUT}}^{N_c}G_{m_{ch}}^2G_{CUT}\times\sum_{n_s=1}^{N_s}\gamma_{n_s}^2\,g_0^2\left(n_s,\frac{f_{s,m_{ch}}+f_{e,m_{ch}}}{2},f_{CUT},f_{CUT}\right)$$

$$\times\frac{1}{4\pi^2\alpha_{0,n_s}\overline{\beta_{2,n_s,XCI}}}\times$$

$$\sum_{j=1}^{2}(-1)^jF_{int}\left(\frac{\pi^2\overline{\beta_{2,n_s,XCI}}}{\alpha_{0,n_s}}\times BW_{CUT}\right.$$

$$\left.\times\left(\frac{f_{s,m_{ch}}+f_{e,m_{ch}}}{2}-f_{CUT}+(-1)^j\frac{BW_{m_{ch}}}{2}\right)\right)$$

Where in (83):

$$\overline{\beta_{2,n_s,SCI}}\triangleq\left(\beta_{2,n_s}+\pi\beta_{3,n_s}\left(2f_{CUT}-2f_{n_s}^{c}\right)\right) \tag{84}$$

$$\overline{\beta_{2,n_s,XCI}}\triangleq\left(\beta_{2,n_s}+\pi\beta_{3,n_s}\left(\frac{f_{s,m_{ch}}+f_{e,m_{ch}}}{2}+f_{CUT}-2f_{n_s}^{c}\right)\right) \tag{85}$$

In [9] the function $F_{int}(.)$ Is approximated as:

$$F_{int}(x)\triangleq j\times\{Li_2(-jx)-Li_2(jx)\}\cong\pi\,\mathrm{asinh}\left(\frac{x}{2}\right) \tag{86}$$

Accepting (86) and noting that $\mathrm{asinh}(.)$ is an odd function, $G_{NLI}^{SCI-XCI}(f_{CUT})$ in (83) can be rewritten as:

$$G_{NLI}^{SCI-XCI}(f_{CUT}) \cong \tag{87}$$

$$\frac{16}{27} G_{CUT}^3 \times \sum_{n_s=1}^{N_s} \gamma_{n_s}^2 \times g_0^2(n_s, f_{CUT}, f_{CUT}, f_{CUT}) \times \frac{asinh\left(\frac{\pi^2 \left|\overline{\beta_{2,n_s,SCI}}\right|}{4\alpha_{0,n_s}} \times BW_{CUT}^2\right)}{4\pi\alpha_{0,n_s}\left|\overline{\beta_{2,n_s,SCI}}\right|}$$

$$+\frac{16}{27} \sum_{\substack{m_{ch}=1 \\ m_{ch}\neq CUT}}^{N_c} G_{m_{ch}}^2 G_{CUT} \times \sum_{n_s=1}^{N_s} \gamma_{n_s}^2\, g_0^2\left(n_s, \frac{f_{s,m_{ch}} + f_{e,m_{ch}}}{2}, f_{CUT}, f_{CUT}\right)$$

$$\times \frac{1}{4\pi\alpha_{0,n_s}\left|\overline{\beta_{2,n_s,XCI}}\right|} \times$$

$$\sum_{j=1}^{2} (-1)^j asinh\left(\frac{\pi^2 \left|\overline{\beta_{2,n_s,XCI}}\right|}{2\alpha_{0,n_s}} \times BW_{CUT}\right.$$

$$\left.\times \left(\frac{f_{s,m_{ch}} + f_{e,m_{ch}}}{2} - f_{CUT} + (-1)^j \frac{BW_{m_{ch}}}{2}\right)\right)$$

Indeed, (87) is composed of two terms added together. First term (red colored) is SCI and second term (blue colored) is XCI.

In [13], an approximated coherency term and in [10],[12] two correction factors are added to (87) as :

$$G_{NLI}^{SCI-XCI}(f_{CUT}) \cong \tag{88}$$

$$\frac{16}{27} G_{CUT}^3 \times \sum_{n_s=1}^{N_s} \gamma_{n_s}^2 \times g_0^2(n_s, f_{CUT}, f_{CUT}, f_{CUT}) \times \rho_{CUT}^{(n_s)}$$

$$\times \frac{asinh\left(\frac{\pi^2 \left|\overline{\beta_{2,n_s,CUT}}\right|}{4\alpha_{0,n_s}} \times BW_{CUT}^2\right) + \rho_{coh} \frac{2 \times Si\left(\pi^2 \left|\overline{\beta_{2,n_s,CUT}}\right| L_s(n_s) \times BW_{CUT}^2\right)}{\pi \times L_s(n_s)\alpha_{0,n_s}} \left[HN(N_s - 1) + \frac{1 - N_s}{N_s}\right]}{4\pi\alpha_{0,n_s}\left|\overline{\beta_{2,n_s,CUT}}\right|}$$

$$+\frac{16}{27} \sum_{\substack{m_{ch}=1 \\ m_{ch}\neq CU}}^{N_c} G_{m_{ch}}^2 G_{CUT} \times \sum_{n_s=1}^{N_s} \gamma_{n_s}^2\, g_0^2\left(n_s, \frac{f_{s,m_{ch}} + f_{e,m_{ch}}}{2}, f_{CUT}, f_{CUT}\right)$$

$$\times \frac{1}{4\pi\alpha_{0,n_s}\left|\overline{\beta_{2,n_s}}\right|} \times \rho_{m_{ch}}^{(n_s)}$$

$$\sum_{j=1}^{2}(-1)^j asinh\left(\frac{\pi^2\left|\overline{\beta_{2,n_s}}\right|}{2\alpha_{0,n_s}} \times BW_{CUT}\right.$$

$$\left.\times\left(\frac{f_{s,m_{ch}}+f_{e,m_{ch}}}{2} - f_{CUT} + (-1)^j\frac{BW_{m_{ch}}}{2}\right)\right)$$

Where in (88), $Si(.)$ and $HN(.)$ are defined as:

$$HN(n) = \sum_{k=1}^{n}\frac{1}{k} \qquad (89)$$

$$Si(x) = \int_{0}^{x}\frac{\sin(t)}{t}dt \qquad (90)$$

In fact, if we set $\rho_{coh} = 0$, $\rho_{CUT}^{(n_s)} = 1$ and $\rho_{m_{ch}}^{(n_s)} = 1$in (88), we will obtain (87) exactly. While in [10], $\rho_{co} = 1$ and $\rho_{CUT}^{(n_s)} \neq 1$ and $\rho_{m_{ch}}^{(n_s)} \neq 1$ and are found through big data approach to improve the accuracy of the formula.

## 7. Big data approch with MCI terms and numerical results

In this section, we add the contribution of MCI term presented in (81) by accepting the approximation presented in (86), to the $G_{NLI}^{SCI-XCI}(f_{CUT})$ presented both in [10] and equation (88) in the previous section. To do this we consider the total NLI as:

$$G_{NLI}(f_{CUT}) \cong \tag{91}$$

$$\frac{16}{27} G_{CUT}^3 \times \sum_{n_s=1}^{N_s} \gamma_{n_s}^2 \times g_0^2(n_s, f_{CUT}, f_{CUT}, f_{CUT}) \times \rho_{CUT}^{(n_s)}$$

$$\times \frac{asinh\left(\frac{\pi^2 \left|\overline{\beta_{2,n_s,SCI}}\right|}{4\alpha_{0,n_s}} \times BW_{CUT}^2\right) + \rho_{coh} \times \frac{2 \times Si\left(\pi^2 \left|\overline{\beta_{2,n_s,SCI}}\right| L_s(n_s) \times BW_{CUT}^2\right)}{\pi \times L_s(n_s)\alpha_{0,n_s}} \left[HN(N_s - 1) + \frac{1 - N_s}{N_s}\right]}{4\pi\alpha_{0,n_s}\left|\overline{\beta_{2,n_s,SCI}}\right|}$$

$$+$$

$$\frac{16}{27} \sum_{\substack{m_{ch}=1 \\ m_{ch} \neq CUT}}^{N_c} G_{m_{ch}}^2 G_{CUT} \times \sum_{n_s=1}^{N_s} \gamma_{n_s}^2\, g_0^2\left(n_s, \frac{f_{s,m_{ch}} + f_{e,m_{ch}}}{2}, f_{CUT}, f_{CUT}\right)$$

$$\times \frac{1}{4\pi\alpha_{0,n_s}\left|\overline{\beta_{2,n_s,XCI}}\right|} \times \rho_{m_{ch}}^{(n_s)} \times$$

$$\sum_{j=1}^{2} (-1)^j asinh\left(\frac{\pi^2 \left|\overline{\beta_{2,n_s,XCI}}\right|}{2\alpha_{0,n_s}} \times BW_{CUT} \times \left(\frac{f_{s,m_{ch}} + f_{e,m_{ch}}}{2} - f_{CUT} + (-1)^j \frac{BW_{m_{ch}}}{2}\right)\right)$$

$$+$$

$$\frac{16}{27} \rho_{MCI} \times \sum_{\substack{m_{ch}=1 \quad n_{ch}=1 \quad k_{ch}=1 \\ (m_{ch}, n_{ch}, k_{ch}) \in S^{MCI}}}^{N_c \quad N_c \quad N_c} G_{m_{ch}} G_{n_{ch}} G_{k_{ch}} \times$$

$$\sum_{n_s=1}^{N_s} \gamma_{n_s}^2 \times g_0^2(n_s, f_1^*, f_2^*, f_{CUT}) \times \frac{1}{16\pi\alpha_{0,n_s}\left|\overline{\beta_{2,n_s,MCI}}\right|}$$

$$\times \sum_{j=1}^{4} (-1)^{(\delta_{j,3} + \delta_{j,4})} asinh\left(\frac{\pi^2 \left|\overline{\beta_{2,n_s,MCI}}\right|}{\alpha_{0,n_s}} \times \left(f_2^* - f_{CUT} + (-1)^j \frac{L_2}{2}\right)\right.$$

$$\left. \times \left(f_1^* - f_{CUT} + (-1)^{(\delta_{j,1} + \delta_{j,4})} \frac{L_1}{2}\right)\right)$$

Where in (91) we have three different terms, summed together, specified with three different colors. The red colored term is SCI, the blue colored term is XCI and the violet colored term is MCI.

In (91), we added $\rho_{MCI}$ and $\rho_{coh}$ to enable us switching on/off the effect of MCI contribution by simply setting $\rho_{MCI}$ to 1 or 0 and also switching on/off the effect of coherent term contribution by simply setting $\rho_{coh}$ to 1 or 0.

In (91), $f_1^* = f_1^*(m_{ch}, n_{ch}, k_{ch})$, $f_2^* = f_2^*(m_{ch}, n_{ch}, k_{ch})$, $L_1 = L_1(m_{ch}, n_{ch}, k_{ch})$ and $L_2 = L_2(m_{ch}, n_{ch}, k_{ch})$ and they are calculated using (7)-(39) replacing $f = f_{CUT}$ . Also $g_0(n_s, f_1^*, f_2^*, f_{CUT})$ in (91) is calculated by replacing $f = f_{CUT}$ in (78). $g_0(n_s, f_{CUT}, f_{CUT}, f_{CUT})$ in (91) is calculated by replacing $f_1^* = f_2^* = f = f_{CUT}$ in (78) and $g_0\left(n_s, \frac{f_{s,m_{ch}}+f_{e,m_{ch}}}{2}, f_{CUT}, f_{CUT}\right)$ by replacing $f_1^* = \frac{f_{s,m_{ch}}+f_{e,m_{ch}}}{2}, f_2^* = f = f_{CUT}$ in (78). Also, $\overline{\beta_{2,n_s,SCI}}$, $\overline{\beta_{2,n_s,XCI}}$ and $\overline{\beta_{2,n_s,MCI}}$ in (91) are define as:

$$\overline{\beta_{2,n_s,SCI}} \triangleq \left(\beta_{2,n_s} + \pi\beta_{3,n_s}\left(2f_{CUT} - 2f_{n_s}^c\right)\right) \quad (92)$$

$$\overline{\beta_{2,n_s,XCI}} \triangleq \left(\beta_{2,n_s} + \pi\beta_{3,n_s}\left(\frac{f_{s,m_{ch}} + f_{e,m_{ch}}}{2} + f_{CUT} - 2f_{n_s}^c\right)\right) \quad (93)$$

$$\overline{\beta_{2,n_s,MCI}} \triangleq \left(\beta_{2,n_s} + \pi\beta_{3,n_s}\left(f_1^* + f_2^* - 2f_{n_s}^c\right)\right) \quad (94)$$

For the reminder we must notice that in the MCI contribution in (91), $(m_{ch}, n_{ch}, k_{ch}) \in S^{MCI}$. $S^{MCI}$ was defined in section (4) but we bring it here again for more emphasis:

$S = \{(m_{ch}, n_{ch}, k_{ch}) \mid \; 1 \le m_{ch} \le N_c\,,\; 1 \le n_{ch} \le N_c\,, 1 \le k_{ch} \le N_c\}$

$S^{SCI} = \{(m_{ch}, n_{ch}, k_{ch}) \in S \mid m_{ch} = n_{ch} = k_{ch} = CUT\}$

$S^{XCI,1} = \{(m_{ch}, n_{ch}, k_{ch}) \in S \mid m_{ch} = n_{ch} = CUT, k_{ch} \ne CUT\}$

$S^{XCI,2} = \{(m_{ch}, n_{ch}, k_{ch}) \in S \mid m_{ch} = k_{ch} = CUT, n_{ch} \ne CUT\}$

$S^{XCI,3} = \{(m_{ch}, n_{ch}, k_{ch}) \in S \mid k_{ch} = n_{ch} = CUT, m_{ch} \ne CUT\}$

$S^{XCI,4} = \{(m_{ch}, n_{ch}, k_{ch}) \in S \mid k_{ch} = n_{ch} \ne CUT, m_{ch} = CUT\}$

$S^{XCI,5} = \{(m_{ch}, n_{ch}, k_{ch}) \in S \mid k_{ch} = m_{ch} \ne CUT, n_{ch} = CUT\}$

$S^{XCI,6} = \{(m_{ch}, n_{ch}, k_{ch}) \in S \mid m_{ch} = n_{ch} \ne CUT, k_{ch} = CUT\}$

$$S^{XCI} = S^{XCI,1} \cup S^{XCI,2} \cup S^{XCI,3} \cup S^{XCI,4} \cup S^{XCI,5} \cup S^{XCI,6}$$
$$= \{(m_{ch}, n_{ch}, k_{ch}) \in S \mid (m_{ch}, n_{ch}, k_{ch}) \in S^{XCI,1} \; or \; (m_{ch}, n_{ch}, k_{ch}) \in S^{XCI,2} \; or \; (m_{ch}, n_{ch}, k_{ch}) \in S^{XCI,3} \; or \; (m_{ch}, n_{ch}, k_{ch}) \in S^{XCI,4} or \; (m_{ch}, n_{ch}, k_{ch}) \in S^{XCI,5} \; or \; (m_{ch}, n_{ch}, k_{ch}) \in S^{XCI,6}\}$$

$$S^{SCI-XCI} = S^{SCI} \cup S^{XCI}$$
$$= \{(m_{ch}, n_{ch}, k_{ch}) \in S \mid (m_{ch}, n_{ch}, k_{ch}) \in S^{SCI} \; or \; (m_{ch}, n_{ch}, k_{ch}) \in S^{XCI}\}$$

$$S^{MCI} = S - S^{SCI-XCI} = S \cap \overline{S^{SCI-XCI}} = \{(m_{ch}, n_{ch}, k_{ch}) \in S \mid (m_{ch}, n_{ch}, k_{ch}) \notin S^{SCI-X} \quad \}$$

To test the accuracy of the (91), we compared its predictions with a benchmark consisting of the full-fledged, numerically integrated EGN-model, in the version [4]. The comparison was run over more than 600 random full C-band (5THz) low-dispersion test systems, which were generated as follows. The WDM comb was centered at 193.41 THz (1550nm). The symbol rate of each channel was randomly chosen among 32, 64, 96 and 128 GBaud with roll-off uniformly-distributed between 0.05 and 0.25. The null-to-null frequency spacing of any two adjacent channels was randomly chosen between 5 and 20 GHz with uniform distribution. The modulation format of each channel was any of PM-QPSK, PM-8QAM, PM-16QAM, PM-32QAM and PM-64QAM. The target OSNRs for max-reach was set to correspond to a GMI of 87% of the entropy (in AWGN). The fiber was DSF (dispersion shifted fiber) with $\alpha = 0.22 \; dB/km$, $\gamma = 1.77 \; (W.Km)^{-1}$ and$\beta_3 = 0.121 \; ps^3/km$. The zero-dispersion wavelength $\lambda_c$ of each span was randomly chosen with a Gaussian distribution with mean 1550nm and std-dev 5nm. The length of each span was randomized and uniformly distributed between 80 and 120km. The EDFAs noise figures were selected randomly between 6 and 7dB. The nominal launch power of each channel was optimized according to the LOGO strategy [3] Eq. (82). The channel under test (CUT) could be anyone out of the five: 1-The channel located at 1550nm (center channel), 2,3- The left and right adjacent neighbors of the center channel in WDM comb and also 4,5- the two extremes in the lowest and highest frequency in WDM comb.

The test procedure was as follows. For each test system, first the max-reach was found using the benchmark EGN-model. Note that due to the great diversity of the randomized links, the max-reach ranged overall between 1 and 16 spans. At max-reach, the quantity $\text{OSNR}_{\text{NL}}$=$P_{\text{ch}}/(P_{\text{ASE}} + P_{\text{NLI}})$ was estimated, both with the benchmark EGN-model, yielding $\text{OSNR}_{\text{NL}}^{\text{EGN}}$, and with the CFM, providing $\text{OSNR}_{\text{NL}}^{\text{CFM}}$. Then the error was assessed as: $ERR = \text{OSNR}_{\text{dB}}^{\text{CFM}} - \text{OSNR}_{\text{dB}}^{\text{EGN}}$ . The quantity ERR is reported in the histograms in the figures.

Figure (4) shows the error histogram when $\rho_{co} \;\; = 0$, $\rho_{MCI} = 0$, $\rho_{SCI} = 1$ and $\rho_{XCI} = 1$. In fact when $\rho_{co} \;\; = 0$, $\rho_{MCI} = 0$, $\rho_{SCI} = 1$ and $\rho_{XCI} = 1$, the CFF in (91) is the same as formula (41) in [3].

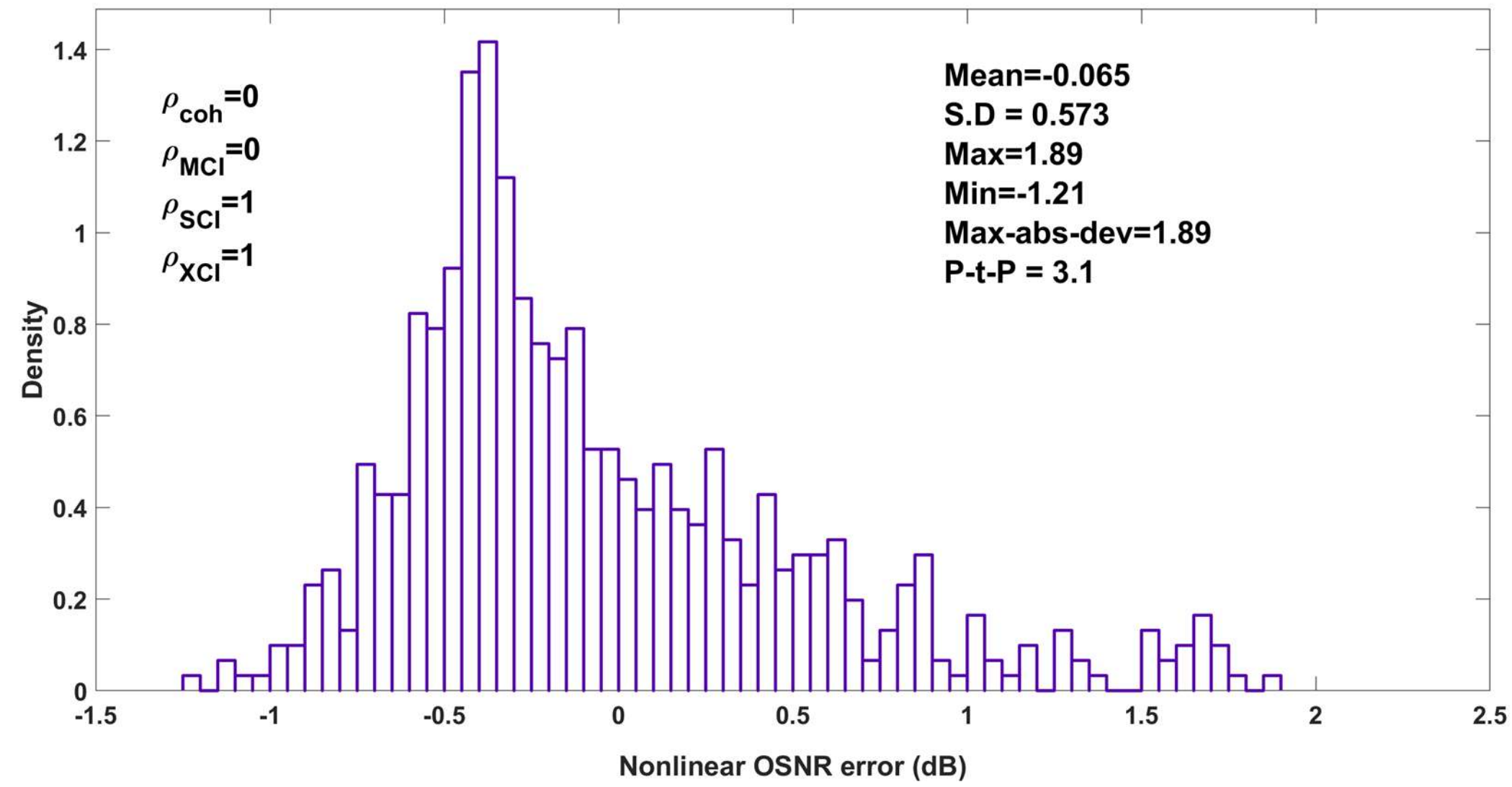

***Figure (4): histogram of the error when $\boldsymbol{\rho_{coh} = 0}$, $\boldsymbol{\rho_{MCI} = 0}$, $\boldsymbol{\rho_{SCI} = 1}$ and $\boldsymbol{\rho_{XCI} = 1}$***

In figure (5), the coherent facto is switched on while MCI is switched off and we have $\rho_{coh} = 1$, $\rho_{MCI} = 0$, $\rho_{SCI} = 1$ and $\rho_{XCI} = 1$.

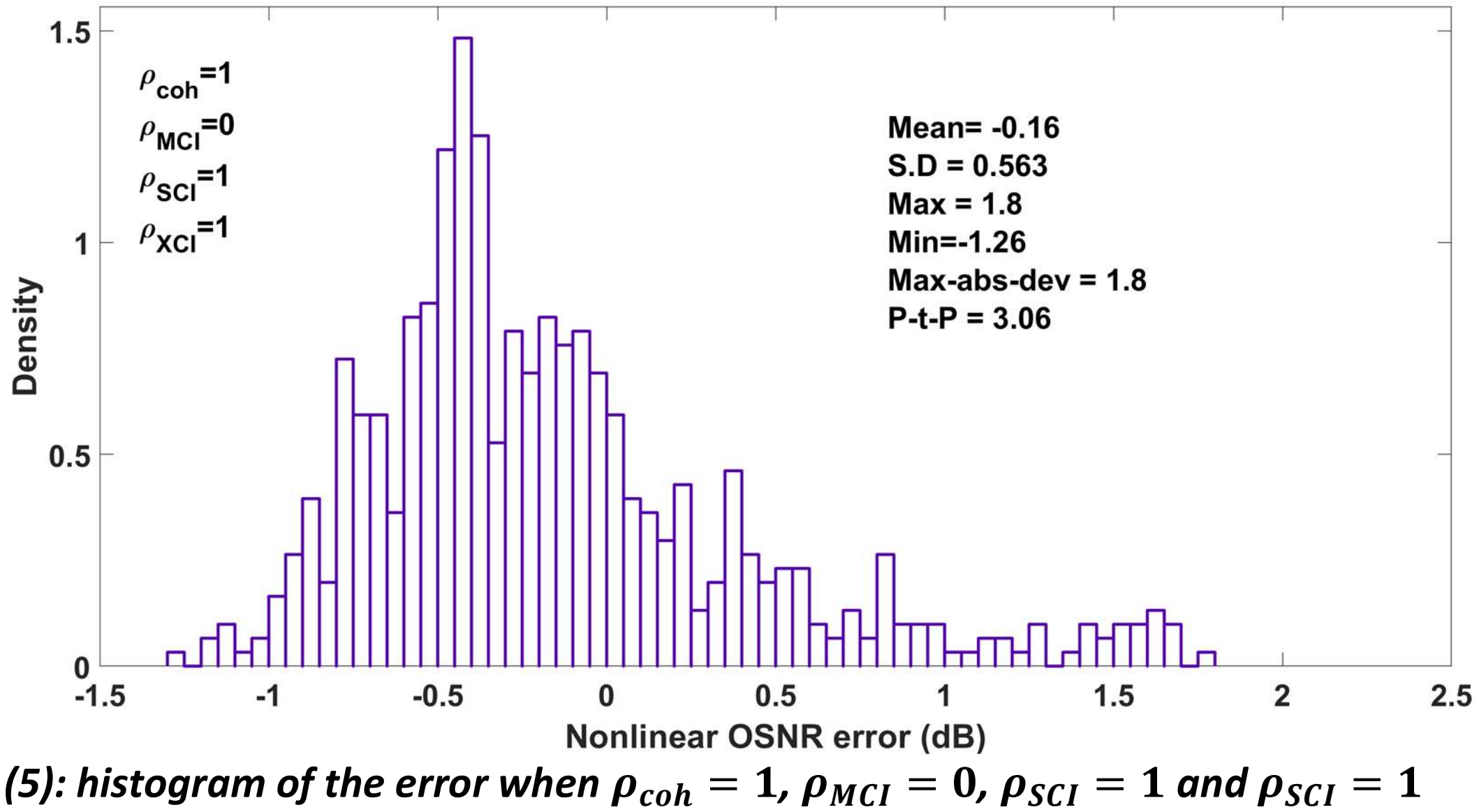

***Figure (5): histogram of the error when $\boldsymbol{\rho_{coh} = 1}$, $\boldsymbol{\rho_{MCI} = 0}$, $\boldsymbol{\rho_{SCI} = 1}$ and $\boldsymbol{\rho_{SCI} = 1}$***

Comparing figure (5)-(6), the coherence term effect is not considerable in the histogram. In figure (6), the MCI contribution is switched on and we have $\rho_{coh} = 1$, $\rho_{MCI} = 1$, $\rho_{SCI} = 1$ and $\rho_{XCI} = 1$.

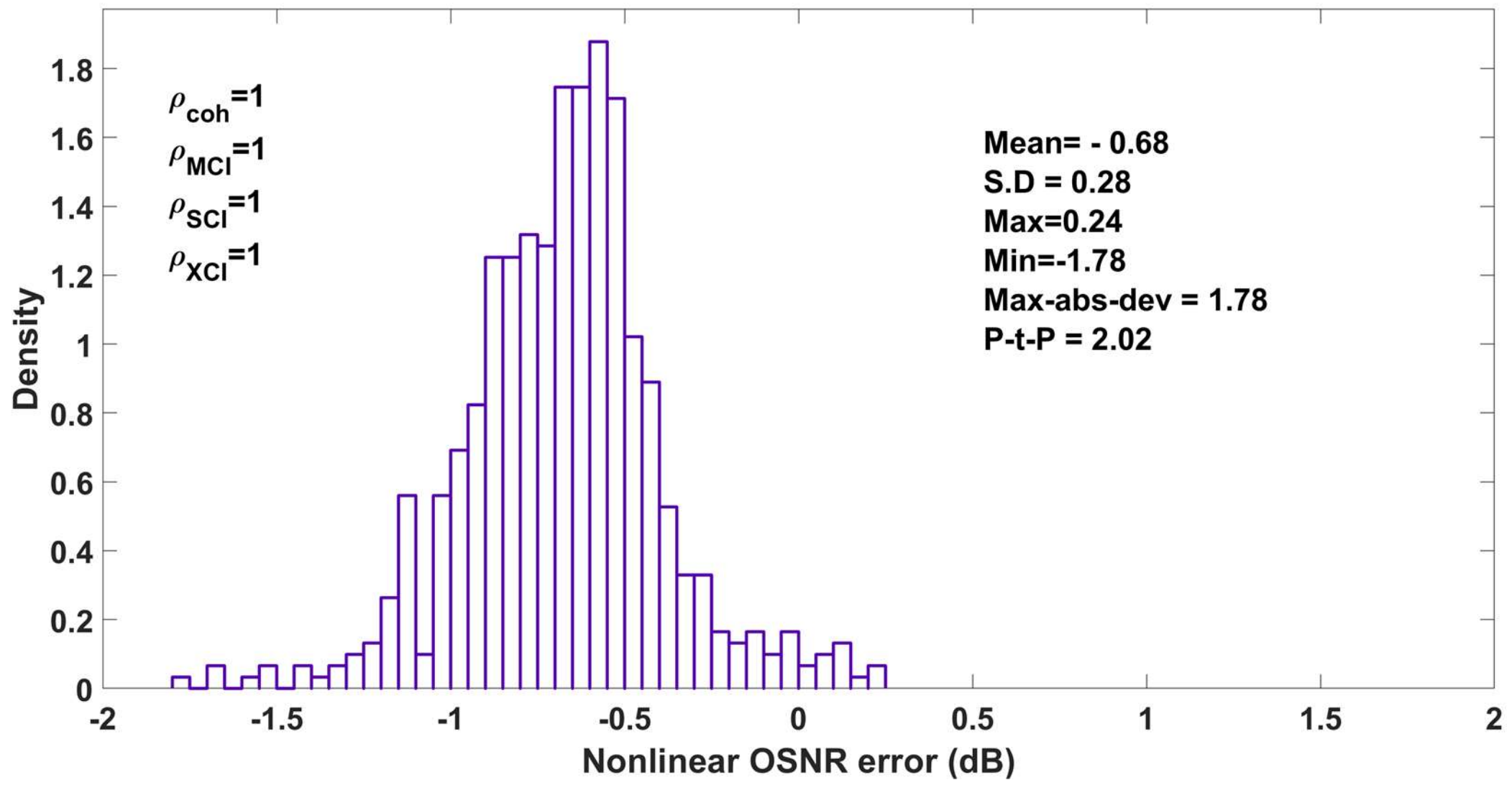

***Figure (6): histogram of the error when $\rho_{coh} = 1$, $\rho_{MCI} = 1$, $\rho_{SCI} = 1$ and $\rho_{XCI} = 1$***

In figure (6), the standard deviation is half of which was in figures (5),(6). However, there is still a nonnegligible mean value (-0.68) available in the figure (6).

To improve accuracy, we then used the correction factors defined in Eq. (95)-(96). They involve the following physical quantities for the CUT and for each channel : the roll-off factor of the CUT channel $r_{CUT}$; the EGN-model format-dependence constant($\Phi_{\text{CUT}}$ for the CUT and $\Phi_{m_{ch}}$for $m_{ch}$'th channel ($m_{ch} \neq CUT$) whose values are listed in [4],[14] based on the channel modulation format); the effective accumulated dispersion $\bar{\beta}_{2,acc,SCI}(n_s, CUT)$at span $n_s$ for CUT and effective accumulated dispersion $\bar{\beta}_{2,acc,XCI}(n_s, m_{ch})$ at span $n_s$ for the $m_{ch}$ 'th channel (their definitions are given in (97)-(98)). Also Kronecker delta function is functioning as an on/off switch in (95)-(96). There are free parameters $a_1$to $a_{23}$. For their best-fitting, we used a standard MSE minimization algorithm on the quantity $ERR$, looking at only 500 out of the 8500+600 (the fitting is done through a mixed and balanced combination of an enlarged set of our previous 8500 test set in [10] and our 600 newly made zero/near zero test set) test-set systems to avoid possible risk of overfitting. The resulting values after fitting optimization for $a_1$to $a_{23}$ are: -0.8509, 1.0923, 0.9305, -0.4097, 0.1652, -15.5857, -0.9648, -0.9826, 0.008273, -0.014253, 253.6104, 0.5174, 0.1695, 0.6250, -1.1281, 0.1591, 0.9497, 0.8592, 0.2265, 0.9047, 0.027842, 0.005731, 1.2457e-41.

$$\rho_{CUT}^{(n_s)} = \left(1 + a_1 r_{CUT}^{a_2}\right)\left(a_3 + a_4 \Phi_{CUT}^{a_5} + a_6\left(1 + a_7 \delta_{\Phi_{CUT},0}\right)\left(1 + a_8\left(BW_{CUT}\right)^{a_9} + a_{10}\log_{10}\left(\left|\overline{\beta}_{2,acc,SCI}(n_s, CUT)\right| + a_{11}\right)\right)\right) \tag{95}$$

$$\rho_{m_{ch}}^{(n_s)} = \left(1 + a_{12} r_{CUT}^{a_{13}}\right)\left(a_{14} + a_{15}\left(\Phi_{m_{ch}} + a_{16}\right)^{a_{17}} + a_{18}\left(\Phi_{m_{ch}} + a_{19}\right)^{a_{20}}\left(1 + a_{21}\delta_{\Phi_{m_{ch}},0}\right)\left(1 + a_{22}\log_{10}\left(\left|\overline{\beta}_{2,acc,XCI}(n_s, m_{ch})\right| + a_{23}\right)\right)\right) \tag{96}$$

$$m_{ch} \neq CUT$$

$$\bar{\beta}_{2,acc,SCI}(n_s, CUT) = \begin{cases} 0 & ; \; for \;\; n_s = 1 \\ \sum_{k=1}^{n_s-1} \overline{\beta_{2,k,SCI}} \times L_s(k) = \sum_{k=1}^{n_s-1} \left\{ \left( \beta_{2,k} + \pi\beta_{3,k}(2f_{CUT} - 2f_k^c) \right) \times L_s(k) \right\} & ; \;\; for \; n_s > 1 \end{cases} \tag{97}$$

$$\bar{\beta}_{2,acc,XCI}(n_s, m_{ch}) = \begin{cases} 0 & ; \; for \; n_s = 1 \\ \sum_{k=1}^{n_s-1} \overline{\beta_{2,k,XCI}} \times L_s(k) = \sum_{k=1}^{n_s-1} \left\{ \left( \beta_{2,k} + \pi\beta_{3,k} \left( \frac{f_{s,m_{ch}} + f_{e,m_{ch}}}{2} + f_{CUT} - 2f_k^c \right) \right) \times L_s(k) \right\} & ; \; for \; n_s > 1 \end{cases} \tag{98}$$

In figure (7), the histogram of error is depicted for $\rho_{coh} = 1, \rho_{MCI} = 1, \rho_{SCI} = eq(95)$ and $\rho_{XCI} = eq(96)$.

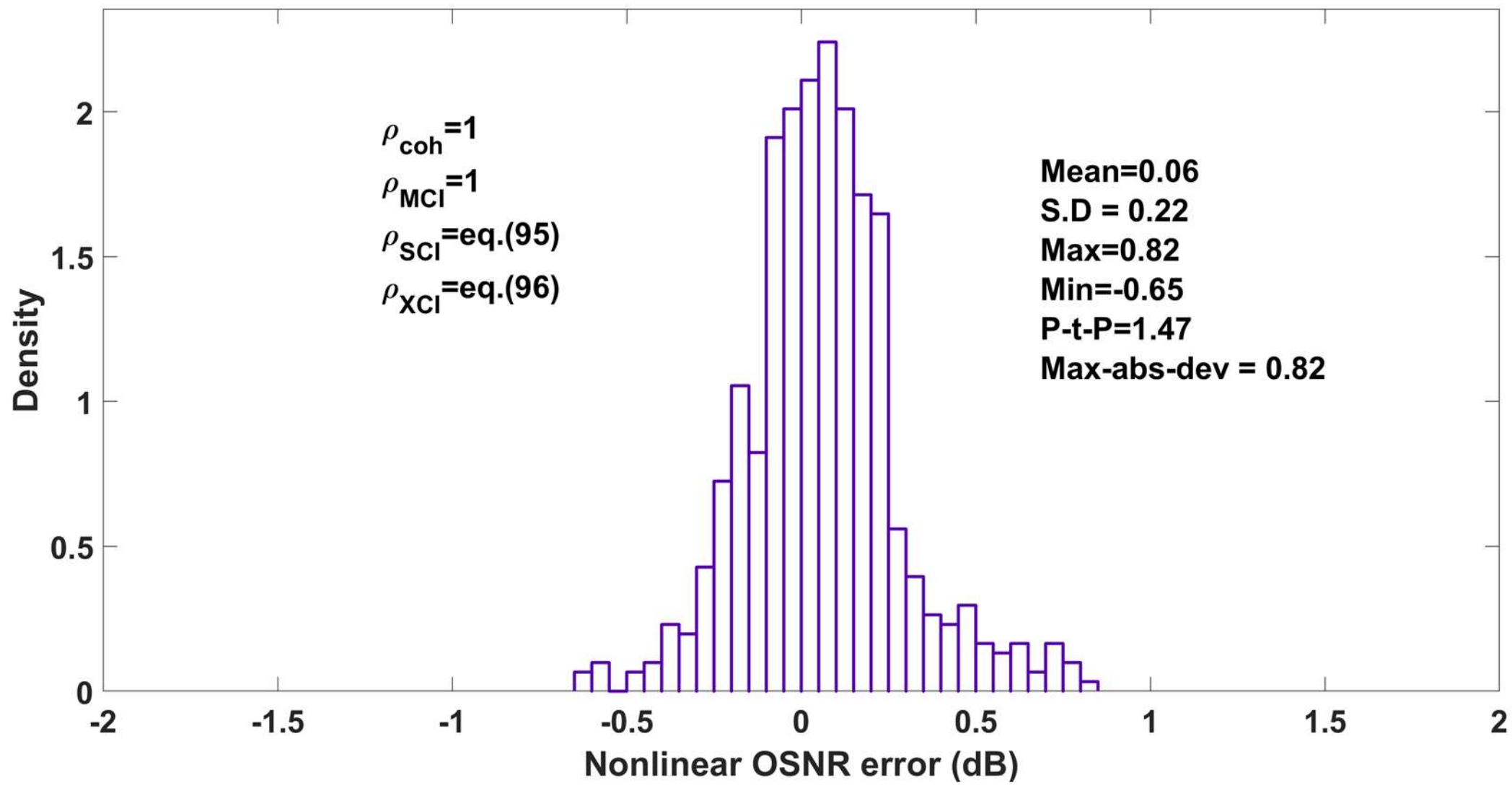

***Figure (7): histogram of the error when $\boldsymbol{\rho_{coh} = 1}$, $\boldsymbol{\rho_{MCI} = 1}$, $\boldsymbol{\rho_{SCI} = eq(95)}$ and $\boldsymbol{\rho_{XCI} = eq(96)}$***

We can see from figure (7) that the accuracy is improved significantly compared to figures (4),(5) while the mean value is shifted to close to the zero value compared to figure (6).

We finally wanted to make sure that by adding the MCI contribution approximation we would not make the model less accurate in dealing with conventional systems. We therefore tested the CFF in (91) with $\rho_{coh} = 1$, $\rho_{MCI} = 1$, $\rho_{SCI} = eq(95)$ and $\rho_{XCI} = eq(96)$ on a 8,500 highly-randomized system test-set, which is an enlarged version of that described in [10]. It includes all QAM modulation formats from QPSK to 256QAM as well as Gaussian-shaped constellations. Three different fiber types are randomly intermixed in the links (SMF, E-LEAF, TWC). The performance of the CFM [10] on this test-set is excellent. When turning on MCI (i.e. going to (91) with $\rho_{co} \ = 1$, $\rho_{MCI} = 1$, $\rho_{SCI} = eq(95)$ and $\rho_{XCI} = eq(96)$ ), the error (ERR) histograms for the lowest, middle and highest frequency channels in the C-band combs are shown in figures (8), (9) and (10) respectively. They are very narrow and comparable to those in [10], so no substantial degradation was induced by adding the approximate MCI terms. Incidentally, the reason why the histogram for the highest-frequency channel figure (10) has worse std-dev than the other two channels, is that the TWC fiber spans possibly present in the links had local dispersion for that channel of only about $D = 0.6\, ps/(nm.km)$, making such channel a near-zero dispersion one. This is why its error histogram in figure (10) looks similar to figure (7).

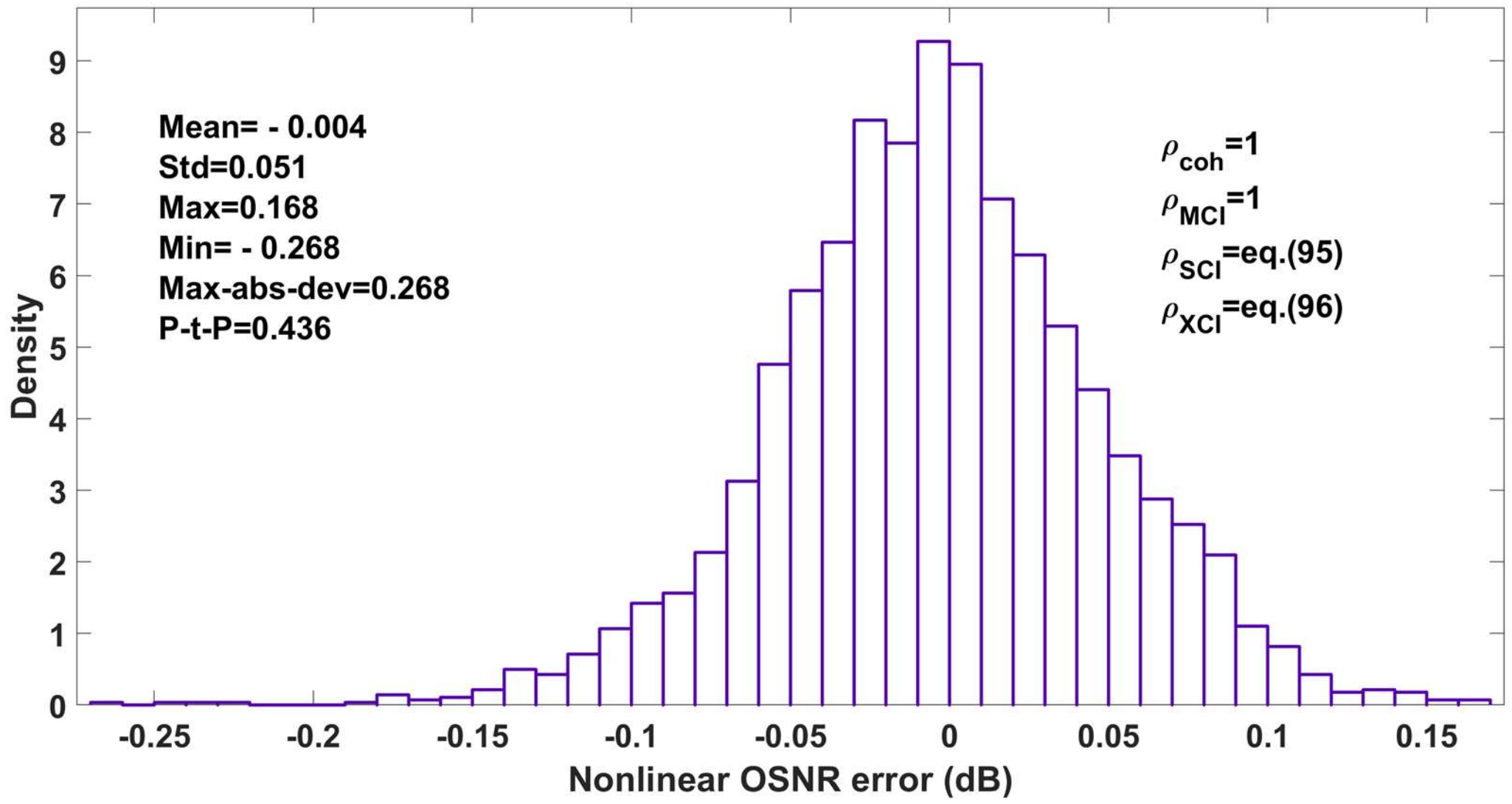

***Figure (8): histogram of the error when $\rho_{coh} = 1$, $\rho_{MCI} = 1$, $\rho_{SCI} = eq(95)$ and $\rho_{XCI} = eq(96)$ for the lowest frequency channel of 8500 test set***

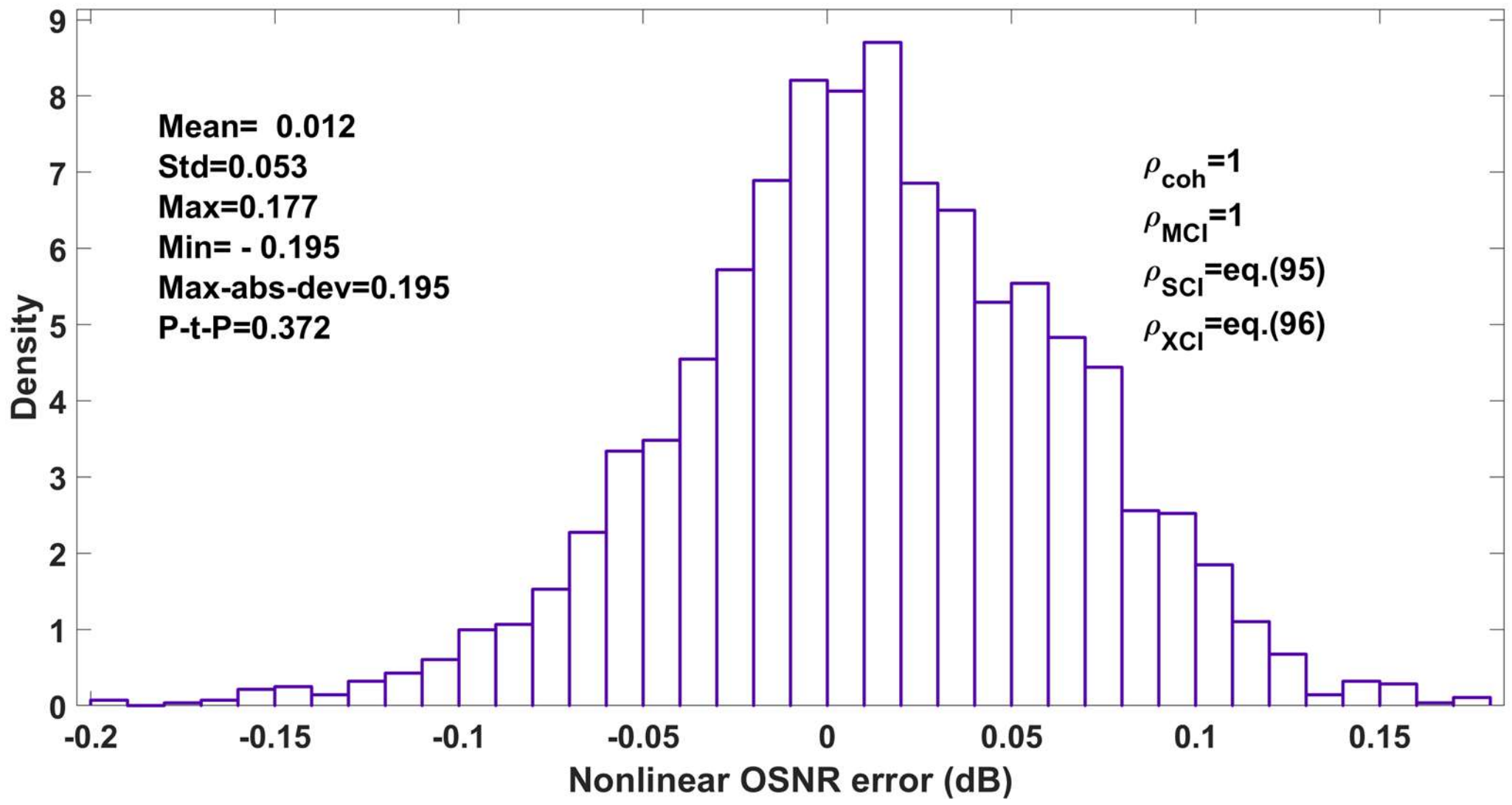

***Figure (9): histogram of the error when $\rho_{coh} = 1$, $\rho_{MCI} = 1$, $\rho_{SCI} = eq(95)$ and $\rho_{XCI} = eq(96)$ for the middle frequency channel of 8500 test set***

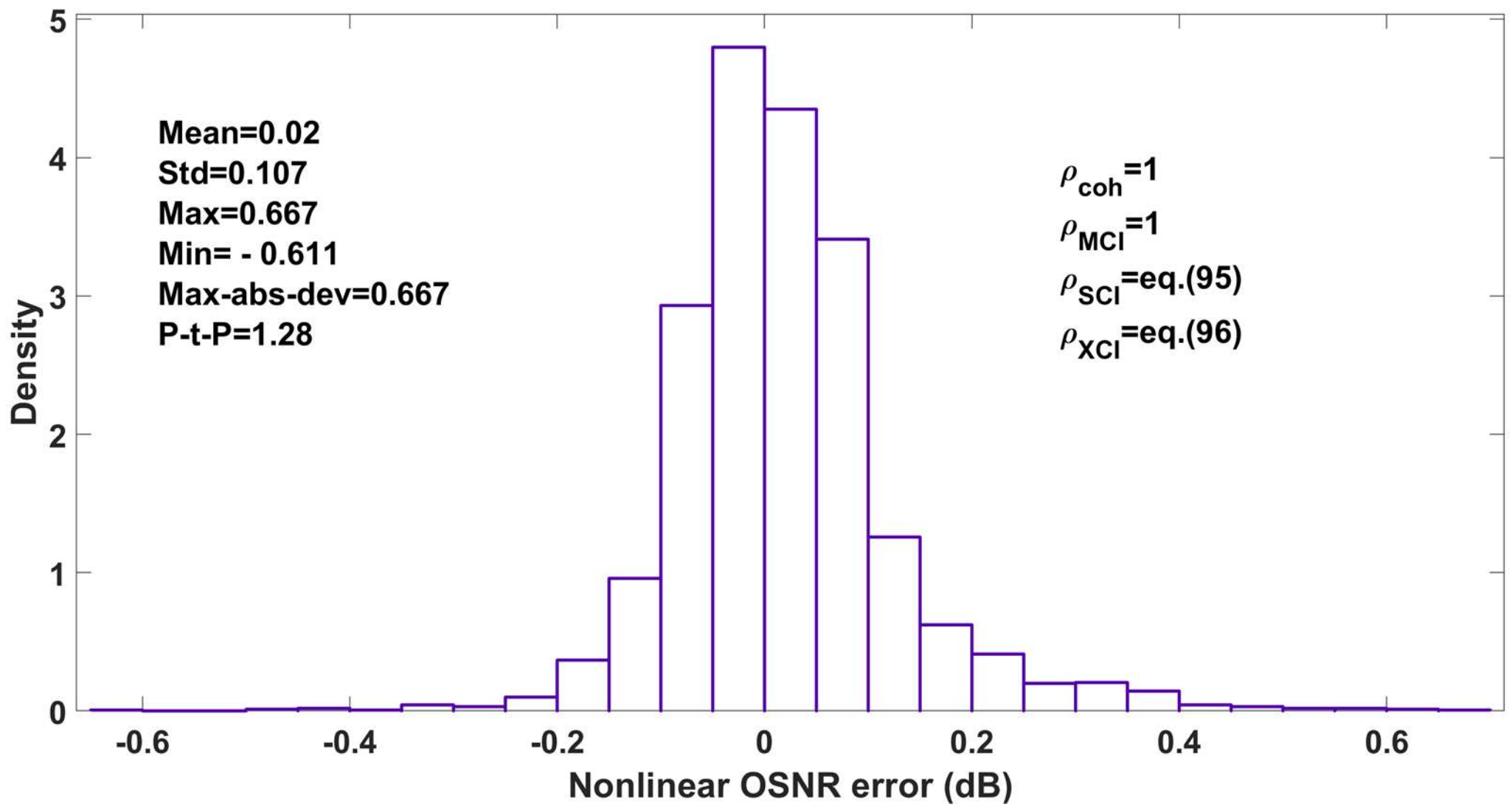

***Figure (10): histogram of the error when $\rho_{coh} = 1$, $\rho_{MCI} = 1$, $\rho_{SCI} = eq(95)$ and $\rho_{XCI} = eq(96)$ for the highest frequency channel of 8500 test set***

## 8. Conclusion

Due to the so-called "capacity crunch", low and even zero-dispersion fibers are being considered for use or re-use. Also, alternative fiber bands are being explored, which can be near or at zero dispersion. In this paper we improve the closed-form model [10] with new analytical terms that make it capable of handling such environments. We finally test it both at low-to-zero dispersion, and over an enlarged version of the general test-set used in [10], for a total of over 9,000 system configurations. The results show it to be a viable tool for real-time management of physical-layer-aware networks even in challenging low-dispersion scenarios.

## 9. Acknowledgemens

This work was supported by Cisco Systems through OPTSYS2020 contract with Politecnico di Torino and by the PhotoNext Center of Politecnico di Torino. Authors would like to thank Stefano Piciaccia and Fabrizio Forghieri from CISCO Photonics for the fruitful discussions and interactions.

## 10. References

[1] A. Mecozzi and R.-J. Essiambre, 'Nonlinear Shannon limit in pseudolinear coherent systems,' JLT, vol. 30, no. 12, pp. 2011–2024, June 15th 2012.

[2] R. Dar, et al., 'Properties of nonlinear noise in long, dispersion-uncompensated fiber links,' Optics Express, vol. 21, no. 22, pp. 25685–25699, Nov. 2013.

[3] P. Poggiolini, et al., 'The GN model of fiber non-linear propagation and its applications,' JLT, vol. 32, no. 4, pp. 694–721, Feb. 2014.

[4] A. Carena, et al., 'EGN model of non-linear fiber propagation,' Optics Express, vol. 22, no. 13, pp. 16335–16362, June 2014.

[5] P. Serena, A. Bononi, 'A Time-Domain Extended Gaussian Noise Model,' JLT, vol. 33, no. 7, pp. 1459–1472, Apr. 2015.

[6] M. Secondini, E. Forestieri, 'Analytical fiber-optic channel model in the presence of cross-phase modulations,' IEEE PTL, vol. 24, no. 22, Nov. 15th, 2012.

[7] R. Dar, et al., 'Pulse collision picture of inter-channel nonlinear interference noise in fiber-optic communications,' JLT, vol. 34, no. 2, pp. 593–607, Jan. 2016.

[8] D. Semrau, R. I. Killey, P. Bayvel, 'A Closed-Form Approximation of the Gaussian Noise Model in the Presence of Inter-Channel Stimulated Raman Scattering,' paper arXiv:1808.07940, Aug. 23rd 2018.

[9] P. Poggiolini, 'A generalized GN-model closed-form formula,' paper arXiv:1810.06545v2, Sept. 24th 2018.

[10] M. Ranjbar Zefreh, A. Carena, F. Forghieri, S. Piciaccia, P. Poggiolini, "A GN/EGN-Model Real-Time Closed-Form Formula Tested Over 7,000 Virtual Links," in Proc. of ECOC 2019, paper W.1.D.3, Dublin, Sep. 2019.

[11] M. Ranjbar Zefreh, P. Poggiolini, "A GN-model closed-form formula considering coherency terms in the Link function and covering all possible islands in 2-D GN integration," arXiv preprint arXiv:1907.09457 (2019)

[12 ] M. Ranjbar Zefreh, et al. "Accurate Closed-Form GN/EGN-Model Formula Leveraging a Large QAM-System Test-Set." IEEE Photonics Technology Letters 31.16 (2019): 1381-1384.

[13] Poggiolini, Pierluigi. "A Closed-Form GN-Model Non-Linear Interference Coherence Term." arXiv:1906.03883 (2019).

[14] P. Poggiolini, Y. Jiang, A. Carena, F. Forghieri 'Analytical Modeling of the Impact of Fiber Non-Linear Propagation on Coherent Systems and Networks,' Chapter 7 in: Enabling Technologies for High Spectral-efficiency Coherent Optical Communication Networks, p. 247-310, Wiley, ISBN: 978-111907828-9, doi: 10.1002/9781119078289.